\documentclass[lettersize,journal]{IEEEtran}
\usepackage{amssymb, amsmath, amsfonts, amsthm}
\usepackage{array}
\usepackage[caption=false,font=normalsize,labelfont=sf,textfont=sf]{subfig}
\usepackage{textcomp}
\usepackage{stfloats}
\usepackage{url}
\usepackage{verbatim}
\usepackage{graphicx}
\usepackage{cite}

\usepackage{graphicx} 
\usepackage{amsmath,amssymb}
\usepackage{multicol}
\usepackage{booktabs}
\usepackage{algorithm}
\usepackage{algpseudocode}
\usepackage[dvipsnames]{xcolor}

\usepackage{bbm, diagbox}

\newcommand{\zd}{\operatorname{zeroDiag}}
\newcommand{\vect}{\operatorname{vec}}

\begin{document}
\title{Optimizing Configuration Selection in Reconfigurable-Antenna MIMO Systems: Physics-Inspired Heuristic Solvers}

\author{Ioannis Krikidis, \IEEEmembership{Fellow, IEEE}, Constantinos Psomas, \IEEEmembership{Senior Member, IEEE}, Abhishek Kumar Singh, \IEEEmembership{Student Member, IEEE}, and Kyle Jamieson, \IEEEmembership{Senior Member, IEEE}   

\thanks{I. Krikidis and C. Psomas are with the Department of Electrical and Computer Engineering, University of Cyprus, Cyprus (email: \{krikidis, psomas\}@ucy.ac.cy).}
\thanks{A. K. Singh and K. Jamieson are with the Department of Computer Science, Princeton University, NJ, USA (email: \{aksingh, kylej\}@princeton.edu).}
\thanks{A preliminary version of this work has been presented at IEEE Int. Conf. Comm. (IEEE ICC'24) \cite{KRI1}}
\thanks{This work has received funding from the European Research Council (ERC) under the European Union's Horizon 2020 research and innovation programme (Grant agreement No. 819819) and the European Union's Horizon Europe programme (ERC, WAVE, Grant agreement No. 101112697). Views and opinions expressed are however those of the author(s) only and do not necessarily reflect those of the European Union or the European Research Council Executive Agency. Neither the European Union nor the granting authority can be held responsible for them.}}

\maketitle

\begin{abstract}
Reconfigurable antenna multiple-input multiple-output (MIMO) is a foundational technology for the continuing evolution of cellular systems, including upcoming 6G communication systems. In this paper, we address the problem of flexible/reconfigurable antenna configuration selection for point-to-point MIMO antenna systems by using physics-inspired heuristics. Firstly, we optimize the antenna configuration to maximize the signal-to-noise ratio (SNR) at the receiver by leveraging two basic heuristic solvers, {\it i.e.,} coherent Ising machines (CIMs), that mimic quantum mechanical dynamics, and quantum annealing (QA), where a real-world QA architecture is considered (D-Wave). A mathematical framework that converts the configuration selection problem into CIM- and QA- compatible unconstrained quadratic formulations is investigated. Numerical and experimental results show that the proposed designs outperform classical counterparts and achieve near-optimal performance (similar to exhaustive search with exponential complexity) while ensuring polynomial complexity. Moreover, we study the optimal antenna configuration that maximizes the end-to-end Shannon capacity. A simulated annealing (SA) heuristic which achieves near-optimal performance through appropriate parameterization is adopted. A modified version of the basic SA that exploits parallel tempering to avoid local maxima is also studied, which provides additional performance gains. Extended numerical studies show that the SA solutions outperform conventional heuristics (which are also developed for comparison purposes), while the employment of the SNR-based solutions is highly sub-optimal.
\end{abstract}


\section{Introduction}

The continuing evolution from 5G to 6G communication systems requires breakthrough physical-layer and networking technologies that can support extremely stringent engineering requirements vis-a-vis capacity, connectivity, and latency. Technologies such as reconfigurable intelligent surfaces, Terahertz communications, semantic communications, fluid antenna systems (FAS), radiated near-field communications, digital twins, \textit{etc.} are just some examples of current research activities to scale up current infrastructure towards 6G \cite{TAT}. All these new communication paradigms significantly increase the computation overhead and demand computing resources with extremely high capabilities. However, classical (silicon) computing architectures cannot be further advanced due to transistors reaching their atomic limits \cite{itrs}. Quantum computing is a promising tool to overcome this computing bottleneck and provide an appropriate computing platform to wireless technologies \cite{KIM}. The application of physics-inspired quantum computing architectures and algorithms in wireless communication systems is a new research area of paramount importance \cite{KIM,huang2023quantum}. 

Quantum computing is built on the fundamental concepts of superposition and entanglement and mainly refers to two basic models \textit{i.e.}, gate-based quantum computing and Ising/annealing model \cite{MCG}. The first model is discrete and uses programmable (reversible) logic gates acting on qubits in a similar fashion to classical digital architectures. By interconnecting basic logic gates, various quantum algorithms can be implemented that provide computation speed-up in comparison to classical counterparts. The work in \cite{BOT} is an informative overview of the application of gate-based quantum algorithms in wireless communication systems. However, gate-based quantum devices are very sensitive to quantum decoherence effects and thus the number of qubits and logic gates that can be applied is limited. The second quantum model (Ising model) is analog and relies on the adiabatic principle of quantum mechanics (Adiabatic theorem \cite{MCG, KIM, YAR}). It is mainly used to solve NP-hard combinatorial optimization problems which are modelled as Ising model instances. By controlling the adiabatic time evolution, the system evolves to a final Hamiltonian whose ground state (lowest energy) encodes the solution of the desired (optimization) problem. Exploiting the adiabatic principles to extract solutions with the lowest energy in a high-dimensional energy landscape, is known as quantum annealing (QA) and is one of many algorithms/systems that are used for optimization/minimization of Ising model instances. 

Ising machines refer to various heuristic solvers designed to find the ground state of the Ising optimization problem ({\it e.g.}, QA \cite{KIM}, coherent Ising machine (CIM) \cite{HAR}, Oscillator-based Ising machine \cite{oim}, \textit{etc.}). While the physical implementation of these systems can vary drastically, ranging from interactions between qubits or optical pulses to coupling between oscillators, all Ising machines take an Ising problem as an input and output a candidate solution. These solvers have been mainly used in the communication literature to solve the maximum likelihood (ML) detection problem for large multi-user multiple-input multiple-output (MU-MIMO) setups. The work in \cite{JAM} introduces the QuAMax MU-MIMO detector, which leverages tools from QA. Specifically, the authors consider the QA D-Wave device, which is commercially available and enables empirical studies at a realistic scale; it is worth noting that state-of-the-art D-Wave architectures are equipped with more than $5,000$ qubits \cite{DWAVE}. In \cite{JAM2}, the authors exploit parallel tempering (ParaMax) to improve the performance of QuAMax. However, these methods rely on a straightforward mapping of the ML MU-MIMO problem to the Ising instance and suffer from an error floor in the bit-error-rate (BER) versus signal-to-noise ratio (SNR) characteristics, and therefore have limited applicability for real systems. To overcome such error floor effects, the work in \cite{ABH2} adopts the CIM solver that uses an artificial optical spin network to find the ground state of the Ising problem; the proposed CIM- based regularized MU-MIMO detector significantly outperforms previous solutions. A more sophisticated CIM technique is proposed in \cite{ABH1}, which converts the original ML MU-MIMO problem into a perturbation correction problem; this technique provides significant performance gains for high-order modulations. Apart from MIMO detection, a few other important problems of interest have been considered in the literature, {\it e.g.,} the authors in \cite{KASI} use the QA D-Wave solver to decode polar codes in wireless cellular networks, while the work in \cite{ROSS} uses similar architectures to design beamforming techniques for reconfigurable intelligent surfaces. Therefore, there has been a rising interest in the application of Ising machines and quantum-inspired computation to complex combinatorial wireless communication problems.

Another physics-inspired meta-heuristic that is based on laws of classical statistical mechanics is simulated annealing (SA) \cite{SA}. The methodology is inspired by the annealing process of materials, in which a solid is heated to a maximum temperature and then cooled-off slowly in a controllable manner until it reaches the desired state with the lowest energy. A Monte Carlo approach that simulates this process was proposed in \cite{metropolis}, which forms the basis of the SA algorithm used in numerous applications \cite{SA}, including communication networks \cite{GZ,ZN,AA,LTN}. Specifically, the authors of \cite{GZ} design a virtual network topology-aware southbound message delivery system, for which they propose two algorithms - a submodular-based approximation algorithm and an SA-based algorithm - for the optimal delivery of southbound messages. In \cite{ZN}, an SA solution is proposed for channel assignment in wireless mesh networks with dynamic spectrum access. The proposed algorithm is evaluated in an experimental setup by using the network simulator 3 (ns3) and its benefits over random channel assignments are presented. The work in \cite{AA} follows a supervised deep learning approach to predict congestion thresholds of cellular LTE base stations. Using these predictions, an SA-based algorithm is designed to minimize the overall congestion of a cluster of base stations; results demonstrate that the SA-based algorithm outperforms existing state-of-the-art tools. An SA-based federated learning scheme is proposed in \cite{LTN}, which lets users keep (probabilistically) their local model instead of adopting the global (average) model. The provided simulations illustrate the gains of the proposed scheme over the conventional federated learning approach.

In this work, we focus on reconfigurable antenna arrays which are an enabling technology for the upcoming 6G communication systems. Reconfigurable/flexible antennas have the capability to modify in a programmable/controllable way their physical and electrical properties ({\it i.e.,} their configuration) to achieve various objectives ({\it e.g.,} increase data rate, control interference, boost SNR, \textit{etc.}). Although the concept is not new \cite{SMI, AFS}, it has recently received a lot of attention due to the recent advances in FAS \cite{WON}. In FAS, the radiated element of the antenna is liquid-based, which moves in a controllable way inside a holder; a reconfiguration, in this case, refers to the alteration of the physical position of the liquid. Most of the work in this area focuses on single antenna setups and studies appropriate channel models and/or signal processing techniques that exploit the liquid dimension \cite{KHA, KRI}. The extension of FAS to MIMO settings is an open problem in the literature; the work in \cite{WON2} adopts an information theoretic standpoint of a ($2$-D) MIMO-FAS, while the associated configuration selection problem is overlooked. 

In this paper, we employ antenna configuration selection to maximize two fundamental performance objectives, {\it i.e.,} the SNR at the receiver's side and the end-to-end Shannon capacity for a point-to-point reconfigurable antenna MIMO system. Since the configuration selection is an NP-hard problem, we adopt physics-inspired meta-heuristics from both classical and quantum statistical mechanics. For the SNR maximization problem, we adopt the CIM solver, which is represented by a set of stochastic differential equations to mimic quantum dynamics, as well as the QA solver, which allows us to embed the equivalent Ising problem into a real-world QA system, {\it i.e.,} D-Wave Advantage. Specifically, the configuration selection problem is firstly formulated as a combinatorial optimization problem with binary variables and multiple constraints. Then, a rigorous mathematical framework that converts the combinatorial problem into an unconstrained quadratic form compatible with CIM and QA implementations is investigated. Numerical results show that the proposed CIM and QA designs outperform classical counterparts and achieve near-optimal performance (similar to exhaustive search (ES)) through appropriate parameterization, while ensuring polynomial complexity. For the capacity maximization problem, we adopt an SA approach which approximates the optimal solution with an annealing and ``cooling'' process, through an appropriate adjustment of the control parameter. A modification of the SA algorithm that exploits parallel tempering is also investigated, which escapes local maxima and gets closer to the optimal solution. The proposed SA-based techniques are compared to conventional heuristics, inspired by the antenna selection problem, which are re-designed for the considered configuration selection problem. Extensive simulation results show that SA-based solutions outperform all conventional solutions while keeping the complexity low. The unique contributions of this study are summarized as follows
\begin{itemize}
\item We design quantum-inspired solvers to optimize the antenna configuration in reconfigurable antenna MIMO systems that maximizes the SNR at the receiver. Specifically, we employ CIM that emulates quantum dynamics and QA where a state-of-the-art QA system (D-Wave Advantage-system4.1) is considered. 
\item A mathematical framework that takes into account the antenna configuration constraints and transforms the original constrained combinatorial problem into CIM- and QA- compatible (unconstrained) quadratic polynomial optimization forms is investigated.
\item We design physics-inspired solvers to optimize the antenna configuration that maximizes the end-to-end information (Shannon) capacity. Specifically, an SA algorithm that achieves near-optimal performance through appropriate parameterization is investigated. A parallel tempering algorithm which runs parallel SA instances and switches between them in time so that to avoid local maxima is also proposed.
\item Numerical studies show that the proposed heuristic solvers achieve optimal performance (similar to ES) and outperform classical counterparts while ensuring polynomial complexity. Particularly, experimental results in the QA D-Wave show the impact of the key parameters, the efficiency of the quantum solver to find the optimal configuration as well as its limitations.  
\end{itemize}
To the best of the authors' knowledge, this is the first time in the literature that physics-based computing tools are used in reconfigurable antenna MIMO systems. Furthermore, we note that the proposed methodologies are generic and not limited to their application in configuration selection but can also be used in any assignment problem that takes a similar form.

The rest of the paper is organized as follows: Section \ref{sy_mo} introduces the system model and presents the combinatorial problems considered. Section \ref{SNR_max} is dedicated to the SNR-maximization problem and presents the two quantum-inspired solvers, {\it i.e.,} CIM and QA. In Section \ref{ca_so}, we deal with the capacity maximization problem and present the SA-based solutions and the associated conventional heuristics. Section \ref{numerical_st} validates the proposed techniques through extensive simulation and experimental results, and Section \ref{concl} summarizes the paper and highlights our primary conclusions. 

\noindent {\it Notation:} Lower and upper case bold symbols denote vectors and matrices, respectively; the superscripts $(\mathbf{U})^{T}$, $(\mathbf{U})^{H}$, $(\mathbf{U})^{-1}$ denote transpose, conjugate transpose and inverse of the matrix $\mathbf{U}$, respectively; $\mathbf{U}\succcurlyeq 0$ means that $\mathbf{U}$ is positive semi-definite, $\det(\mathbf{U})$ denotes the determinant of the matrix $\mathbf{U}$, $\mathbf{I}$ denotes an identity matrix of appropriate dimension, $\mathbf{0}_{k\times n}$ ($\mathbf{0}_{k}$) and $\mathbbm{1}_{k\times n}$ ($\mathbbm{1}_{k}$) denote a $k \times n$ null and an all-ones matrix ($k\times 1$ column vector), respectively; $\mathbbm{C}^{k\times n}$ denotes the space of $k\times n$ matrices with complex entries, $\mathcal{CN}(\mu,\sigma^2)$ represents the complex Gaussian distribution with mean $\mu$ and variance $\sigma^2$; $\textrm{Tr}(\cdot)$ is the trace operator, $\mod$ denotes the modulo operator, and $\lceil \cdot \rceil$ rounds up toward positive infinity; $\| \cdot \|_{\infty}$ denotes the $\infty$-norm ($\max$-norm), $\vect(\mathbf{U})$ converts a matrix into a column vector, $\zd(\mathbf{U})$ sets the diagonal elements to zero, $\mathbbm{E}(\cdot)$ denotes the statistical expectation; $|z|$ is the absolute value of a complex scalar $z$, and the $\log(\cdot)$ function has base-$2$ by default.

\section{System model}\label{sy_mo}
We consider a fundamental point-to-point MIMO setup consisting of $N_T$ and $N_R$ antennas at the transmitter and the receiver, respectively. Each antenna is reconfigurable and can change its physical and electrical properties in a controllable way; $N$ distinct states are assumed at each antenna. A potential implementation of this setup refers to MIMO-FAS, where the liquid in each antenna can be displaced to $N$ predefined locations (ports) \cite{WON}. However, the model considered holds for any type of reconfigurable antenna MIMO. Moreover, we assume that there exists full channel state information knowledge. Even though channel estimation is a challenging task, due to the large number of channels, efficient and accurate techniques have been proposed, based on machine learning \cite{10018377} and compressed sensing \cite{10236898}.

\begin{figure}
\includegraphics[width=\linewidth]{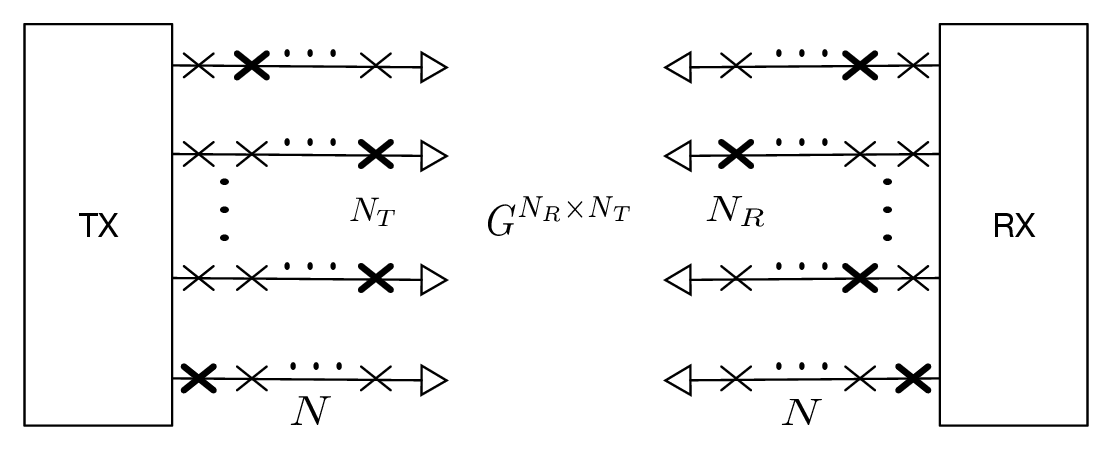}
\caption{The considered point-to-point reconfigurable MIMO system with $N_T$ and $N_R$ antennas at the transmitter and the receiver, respectively; $N$ states at each antenna. The symbol $\times$ represents an antenna state, while the set of bold symbols $\pmb{\times}$ corresponds to the selected configuration.}\label{system}
\end{figure}

To facilitate the mathematical formulation, we introduce the complete MIMO channel matrix $\mathbf{G}^{NN_R\times N N_T}$, whose entries correspond to the channels between the transmit and and receive antennas for all possible antenna states, {\it i.e.}, $g_{i,j}$ is the channel coefficient between the transmit antenna $\lceil j/N \rceil$ with state $(j\mod N)$ and the receive antenna $\lceil i/N \rceil$ with state $(i\mod N)$. Without loss of generality, we assume mutual independence between the antenna states with Rayleigh block fading channels, {\it i.e.}, $g_{i,j}\sim \mathcal{CN}(0,1)$ \cite{SMI}. According to the principles of reconfigurable antennas, only one state per antenna can be active in each operation time. Therefore, a configuration of the considered MIMO setup refers to a selection of specific $N_T$ and $N_R$ antenna states at the transmitter and receiver, respectively. Subsequently, a configuration selection reduces the complete channel matrix into the conventional $N_R \times N_T$ MIMO matrix. Fig. \ref{system} schematically depicts the system model.  

The diagonal matrices $\mathbf{X}^{N N_T\times N N_T}$ and $\mathbf{Y}^{N N_R\times N N_R}$ are introduced to enable configuration selection at the transmitter and the receiver side, respectively. The elements of these matrices are binary, that is, $x_{i,i},y_{i,i} \in \{0,1\}$, where a value equal to one (or zero) means that the state ($i\mod N$) of the antenna $\lceil i/N \rceil$ is selected (or not selected). Hence, the baseband representation of the MIMO link can be written as 
\begin{align}
\mathbf{y}=\Phi(\mathbf{H})\mathbf{x}+\mathbf{z},
\end{align}
where $\mathbf{H}=\mathbf{Y}\mathbf{G}\mathbf{X}$ is the equivalent channel matrix incorporating the configuration selection at both sides, the function $\Phi(\cdot)$ reduces the dimension of its matrix argument by removing all-zero columns/rows, $\mathbf{y} \in \mathbbm{C}^{N_R\times 1}$ is the received signal vector, $\mathbf{x} \in \mathbbm{C}^{N_T\times 1}$ is the transmitted signal vector with covariance matrix $\mathbf{S}=\mathbbm{E}(\mathbf{x}\mathbf{x}^H)$ and $\textrm{Tr}(\mathbf{S})\leq P$, where $P$ is the transmit power, and $\mathbf{z}$ is the additive white Gaussian noise vector with elements $z_{i}\sim \mathcal{CN}(0,1)$.

In this work, we aim to maximize two fundamental performance objective functions for the reconfigurable antenna MIMO system: the SNR at the receiver\footnote{Maximizing the SNR at the receiver is equivalent to maximizing the energy harvesting at the receiver in wireless power transfer MIMO setups when a linear energy harvesting model is considered \cite{ZHA}.} (problem (P1)), and the end-to-end Shannon capacity (problem (P2)). The associated combinatorial optimization problems are given as follows
\begin{subequations}\label{opt11}
\begin{align}
\mathbf{(\text{P1})} &\max_{\mathbf{X},\mathbf{Y},\mathbf{S}}\;\;Q\equiv \textrm{Tr}\{\Phi(\mathbf{H}) \mathbf{S}\Phi(\mathbf{H})^H\} \nonumber \\
&\;\;\;\;\;\;\;\;\;\;\;\;\;\;\,=\textrm{Tr}(\mathbf{X}\mathbf{G}^{H}\mathbf{Y\mathbf{S}YGX})\\
&\textrm{subject to}\;\sum_{i=kN+1}^{(k+1)N} x_{i,i}=1,\quad k=0,\ldots,N_{T}-1, \label{c1}  \\
&\;\;\;\;\;\;\;\;\;\;\;\;\;\;\;\;\sum_{i=k N+1}^{(k+1)N} y_{i,i}=1,\quad k=0,\ldots,N_{R}-1, \label{c2} \\
&\;\;\;\;\;\;\;\;\;\;\;\;\;\;\;\;\;\textrm{Tr}(\mathbf{S})\leq P,\;\;\;\;\mathbf{S}\succcurlyeq 0. \label{c3}
\end{align}
\end{subequations}
The constraints in \eqref{c1} and \eqref{c2} correspond to the physical limitation that only one state is active at each antenna for both the transmitter and the receiver, respectively. In a similar way, the capacity maximization problem is written as
\begin{subequations}
\begin{align}
\mathbf{(\text{P2})} &\max_{\mathbf{X},\mathbf{Y},\mathbf{S}} \;\; C\equiv \log\det (\mathbf{I}+\Phi(\mathbf{H})\mathbf{S}\Phi(\mathbf{H})^H)\\
&\textrm{subject to}\;\eqref{c1}, \eqref{c2}, \eqref{c3}.
\end{align}
\end{subequations}
The considered optimization problems have both binary (antenna state selection) variables and continuous variables (power allocation) and are of combinatorial nature; they are both NP-hard problems with exponential complexity and their optimal/joint solution mainly requires an exchaustive searching (ES) over all configurations while the optimal power control is computed per configuration. By taking into account the structure of the problems (P1) and (P2), we decouple the initial formulations into two independent sequential sub-problems, {\it i.e.,} (i) {\it Antenna configuration sub-problem}, where the optimal configuration is computed by assuming a symmetric power allocation with $\mathbf{S}=\frac{P}{N_T}\mathbf{I}$, and (ii) {\it Power allocation sub-problem}, where the optimal power control is computed by considering the configuration given by the antenna configuration sub-problem.  

The second sub-problem is well known in the literature and can be solved by using standard convex optimization tools. Specifically, for a given antenna configuration, the optimal power allocation is given by the water-filling power allocation solution for the capacity maximization problem \cite[eq. (3)]{ZHA}. On the other hand, the covariance matrix that maximizes SNR is equal to $\mathbf{S}_{\rm SNR} = P\mathbf{v}_1\mathbf{v}_1^H$, where $\mathbf{v}_1$ denotes the first column of the matrix $\mathbf{V}$, given by the singular value decomposition $\Phi(\mathbf{H}) = \mathbf{V}\mathbf{\Sigma}\mathbf{U}^H$; $\mathbf{V}$ and $\mathbf{U}$ are complex unitary matrices of appropriate dimension, and $\mathbf{\Sigma}$ is a diagonal matrix containing the singular values of the matrix in descending order \cite[prop. 2.1]{ZHA}. 

\begin{figure}\centering
\includegraphics[width=\linewidth]{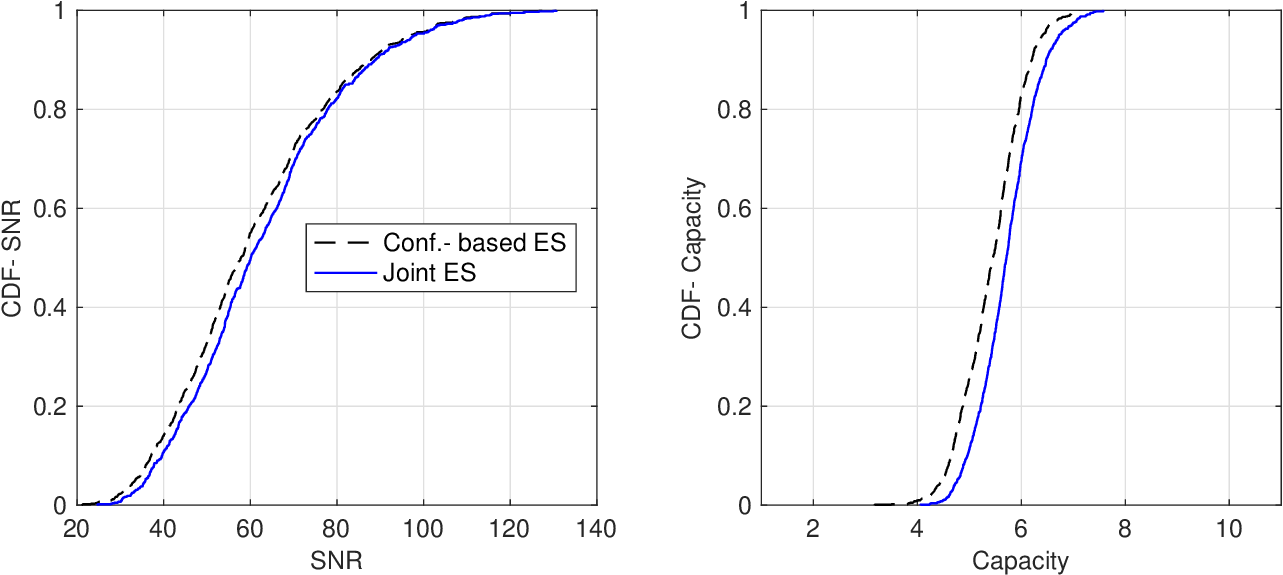}
\caption{(left) CDF of the SNR objective function; (right) CDF of the capacity objective function, for both joint and configuration-based ES schemes. Setup with $N_T=N_R=2$, $N=2$ and $P=10$ (dB).}\label{fig:decoupling}
\end{figure}

To demonstrate the effectiveness of the problem decomposition, Fig. \ref{fig:decoupling} shows the cumulative distribution function (CDF) of the considered objective functions, ({\it i.e.,} SNR and Shannon capacity) for two different ES schemes. Specifically, we demonstrate the CDF performance for (i) a full/joint ES that solves the power allocation problem for each antenna configuration, and (ii) a decoupled (configuration-based) ES that computes the optimal configuration, given a symmetric power allocation, while the power allocation problem is solved only for the selected configuration. The numerical investigations show that the CDF curves are very close for both objectives, which indicates that the decomposition provides an efficient (albeit sub-optimal) solution, at least for setups of practical interest; it is worth noting that a similar decoupling is used in \cite{WON2}. Furthermore, since the second sub-problem (the power allocation) is well-studied in the literature with an optimal closed-form solution available, in this work, we focus on the antenna configuration sub-problem, which is an interesting and challenging problem on its own.

\subsection{Conventional solutions and benchmarks}
For the antenna configuration sub-problem, we consider three conventional techniques that are used as performance benchmarks.

\subsubsection{Exhaustive search (ES)} The ES is the optimal solution and computes the objective function for all possible combinations of the antenna states. The algorithm requires $N^{N_T} \times N^{N_R}=N^{N_T+N_R}$ calculations of the objective function (one for each combination) and therefore its complexity becomes exponential with the number of antennas/states, {\it i.e.,} $\mathcal{O}(N^{N_T+N_R})$; it is prohibited for MIMO setups with high number of antennas/configurations.

\subsubsection{Norm-based selection algorithm (NSA) \cite{SMI}} In the NSA, the receiver and the transmitter select the configurations corresponding to the highest/strongest row and column norms (Euclidean), respectively. More specifically, the selection is firstly performed at one end of the link (\textit{e.g.}, the receiver) and each antenna selects the configuration with the highest row norm. By using the selected row configurations, each antenna at the other end of the link (\textit{e.g.}, the transmitter) selects the configuration with the highest column norm. This selection scheme reduces significantly the number of computations, {\it i.e.,}, $\mathcal{O}(N(N_T+N_R))$ and thus it has low complexity and high practical interest. 

\subsubsection{Random selection (RS)} The RS is a simple scheme where a random state is selected at each antenna. It does not require complicated computations or any intelligence, {\it i.e.,} $\mathcal{O}(1) $. 

\section{SNR maximization with CIM and QA solvers}\label{SNR_max}
In this section, we consider the SNR objective function and study the optimal antenna configuration that maximizes the SNR at the receiver. By assuming a symmetric power allocation, the original problem $(\textrm{P1})$ is simplified to   
\begin{subequations}\label{opt33}
\begin{align}
\mathbf{(\text{P3})}\; &\max_{\mathbf{X},\mathbf{Y}}\;Q_0 \equiv \textrm{Tr}\{\Phi(\mathbf{H})\Phi(\mathbf{H})^H\} \nonumber \\
&\;\;\;\;\;\;\;\;\;\;\;\;\;\,=\textrm{Tr}(\mathbf{X}\mathbf{G}^{H}\mathbf{YG})\\
&\textrm{subject to}\;\;\eqref{c1}, \eqref{c2},
\end{align}
\end{subequations}
where the power allocation (constant) term $P/N_T$ is omitted from the objective function for simplicity, since it does not affect the solution. To solve ($\textrm{P3}$), we consider two heuristic solvers, {\it i.e.,} CIMs which exploit quantum mechanical dynamics and mimic QA designs and QA where real-world experiments in a state-of-the-art D-Wave device are considered. 

\subsection{CIM solver}

To apply the CIM heuristic solver for the problem ($\textrm{P3}$), we convert it to an instance of the Ising problem. We firstly highlight the basic properties of the CIM as well as the associated system of stochastic differential equations. Then, the transformation of the considered optimization problem into CIM compatible form is presented.  

\subsubsection{CIM preliminaries}
A CIM is a heuristic solver for finding the ground state of an Ising optimization problem, which is a quadratic binary optimization problem and can be expressed as
\begin{equation}
     \arg \min_{\forall i,s_i\in \{-1,+1\}}-\sum_{i \neq j}J_{ij}s_{i}s_{j},
\end{equation}
or equivalently in a vector form given by
\begin{equation}
    \arg \min_{\mathbf{s}\in \{-1,+1\}^N}  - \mathbf{s}^T\mathbf{J}\mathbf{s},
\end{equation}
where $s_i$ are the spin variables taking values from $\{-1,+1 \}$ and $J_{ij}$ are the coefficients of the Ising problem being solved. CIMs were designed to utilize an artificial optical spin network \cite{marandi2014network,mcmahon2016fully,inagaki2016coherent} to solve Ising optimization problems. The dynamics of such systems can be approximately modeled as \cite{ahc}
\begin{equation}
    \forall i\text{,      }\dfrac{dx_i}{dt} = (1-p)x_i - x_i^3 + \epsilon \sum_{j\neq i} J_{ij}x_j, 
\end{equation}
where $p$ and $\epsilon$ are system parameters (constants) and $x_i$ are state variables describing such systems. While CIMs are designed to find the global optimal, they can get stuck in local minima and limit cycles \cite{ahc}. An enhanced model with amplitude heterogeneity correction (AHC) \cite{ahc} that destabilizes these local minima can be used to improve the overall performance. An AHC-based CIM model can be described as \cite{ahc}
\begin{equation}
     \forall i \text{,      }\dfrac{dx_i}{dt} = (1-p)x_i - x_i^3 + \epsilon e_i \sum_{j\neq i} J_{i,j}x_j,
     \label{eq:cim1}
\end{equation}
\begin{equation}
     \forall i \text{,      } \dfrac{de_i}{dt} = -\beta(x_i^2 - a)e_i,\text{    }e_i > 0,
     \label{eq:cim2}
\end{equation}
where $\beta$ and $a$ are constant system parameters and $x_i$ and $e_i$ are the state variables of the system. The spin solution corresponding to a CIM state is simply given by $s_i = \operatorname{sgn}(x_i)$, where $\operatorname{sgn}(\cdot)$ denotes the sign function. In this work, we simulate an AHC-based CIM model by performing numerical integration of (\ref{eq:cim1}) and (\ref{eq:cim2}) for $1,000$ time-steps with $dt = 0.01$. Based on empirical evaluation, we set the parameters of the model to $p=0.98$, $\beta = 1$, $a = 2$, and $\epsilon = \gamma t$, where $t$ is the time at a specific time-step and $\gamma = 100$ \cite{ABH1}.

\subsubsection{CIM formulation and design}\label{sec_cim}

We now present a mathematical framework that transforms the original combinatorial optimization problem in \eqref{opt33} to the Ising model and specifically in an appropriate form that is compatible with  CIM optical hardware implementations. 

Firstly, we introduce a binary vector $\mathbf{b}$ that integrates the diagonal elements from both $\mathbf{X}$ and $\mathbf{Y}$ matrices, {\it i.e.},
\begin{align}
\mathbf{b}^T&=\left[x_{1,1},\ldots,x_{NN_{T},NN_T},y_{1,1},\ldots,y_{NN_{R},NN_R}\right] \nonumber \\
&=\left[b_{1},\ldots,b_{N(N_{T}+N_{R})}\right],\;\;\; b_{i}\in\left\{0,1 \right\}.
\end{align}

We also define the symmetric matrix $\mathbf{Q}$ of dimension $N(N_T+N_R)\times N(N_T+N_R)$ given by 
\begin{align}\label{Q_matrix}
\mathbf{Q}=\left[ \begin{matrix}
\mathbf{0}_{NN_T\times NN_T} & \frac{1}{2}\mathbf{T} \\[0.5ex] 
\frac{1}{2}\mathbf{T}^T & \mathbf{0}_{NN_R\times NN_R}\\[0.5ex]
\end{matrix}\right],
\end{align} 
\noindent where the matrix $\mathbf{T}^{NN_T\times N N_R}$ consists of the squared amplitudes of the channel coefficients, \textit{i.e.},
\begin{align}
\mathbf{T}=\left[ \begin{matrix}
\left| g_{1,1}\right| ^{2} & \left| g_{2,1}\right| ^{2} & \cdots & \left| g_{NN_{R},1}\right| ^{2}\\
\vdots & \vdots & \ddots & \vdots\\
\left| g_{1,NN_{T}}\right| ^{2} & \left| g_{2,NN_{T}}\right| ^{2} & \cdots & \left| g_{NN_{R},NN_{T}}\right| ^{2}\\[0.5ex] 
\end{matrix}\right].
\end{align}
By using the above definitions, the original optimization problem in \eqref{opt33} can be converted to the following quadratic form 
\begin{subequations}\label{opt22}
\begin{align}
&\arg\max_{\mathbf{b}}\;\mathbf{b}^T\mathbf{Q}\mathbf{b} \\
&\textrm{subject to}\; P_k(\mathbf{b})=\sum_{i=k N+1}^{(k+1)N}b_i-1=0,\\
&\;\;\;\;\;\;\;\;\;\;\;\;\;\;\;\;k=0,\ldots,N_{T}+N_{R}-1.\nonumber
\end{align}
\end{subequations}

The above quadratic optimization problem has constraints on the binary variables and therefore can not be used directly for quantum implementations. To overcome this bottleneck, we consider the following equivalent quadratic representation of the $k^{th}$ constraint 
\begin{align}
P_k(\mathbf{b})=\left( \sum_{i=kN+1}^{(k+1)N}b_i-1\right) ^{2}=\left( \sum_{i}b_i\right) ^{2}-2\sum_{i}b_i+1. \label{k-th}
\end{align}

By ignoring the constant terms in \eqref{k-th}, the $k^{\textrm{th}}$ constraint can be written in quadratic binary form as follows
\begin{align}
P_k(\mathbf{b})=\mathbf{b}^{T}\mathbf{A}_{k}\mathbf{b}-2\mathbf{h}_k^{T}\mathbf{b}, 
\end{align}
where $\mathbf{h}_k^{T}=\left[\mathbf{0}_{kN} \quad \mathbbm{1}_{N} \quad \mathbf{0}_{(N_{T}+N_R-k-1)N}\right]^T$ and $\mathbf{A}_k$ is defined in \eqref{Ak}.

\begin{figure*}[ht]\vspace{-4mm}		
\begin{align}
\mathbf{A}_{k}=\left[\begin{array}{ccc}
\mathbf{0}_{kN\times kN} & \mathbf{0}_{kN\times N} & \mathbf{0}_{kN\times (N_{T}+N_R-k-1)N}\\
\mathbf{0}_{N\times kN} & \mathbbm{1}_{N\times N} & \mathbf{0}_{N\times (N_{T}+N_R-k-1)N}\\
\mathbf{0}_{(N_{T}+N_R-k-1)N\times kN} & \mathbf{0}_{(N_{T}+N_R-k-1)N\times N} & \mathbf{0}_{(N_{T}+N_R-k-1)N\times (N_{T}+N_R-k-1)N}\\
\end{array} \right]. \label{Ak}
\end{align}
\noindent\makebox[\linewidth]{\rule{18.3cm}{.4pt}}\vspace{-2mm}
\end{figure*}

Since all the constraints have the same impact on the problem considered, we combine them in a single aggregate constraint through summation. More specifically, we have 
\begin{align}
P_0(\mathbf{b})&=\sum_{k=0}^{N_T+N_R-1}P_k(\mathbf{b}) \nonumber \\
&=\mathbf{b}^T \left(\sum_{k=0}^{N_T+N_R-1} \mathbf{A}_k \right)\mathbf{b}-2 \left(\sum_{k=0}^{N_T+N_R-1}\mathbf{h}_k^T \right)\mathbf{b} \nonumber\\
&=\mathbf{b}^T \mathbf{R}\mathbf{b}-2\mathbbm{1}_{N(N_T+N_R)}^T \mathbf{b}. \label{quad_const}
\end{align} 
The next step of the mathematical framework is to convert the binary vector $\mathbf{b}$ into the spin vector $\mathbf{s}$, where the spin variables $s_i$ take values in $\{-1,+1\}$. By considering the transformation $b_i=\frac{1}{2}(s_i+1)$ \cite{KIM}, the expression in \eqref{quad_const} is equivalent to 
\begin{align}
P_0(\mathbf{s})&=\frac{1}{4}\mathbf{s}^T \mathbf{R}\mathbf{s}+\left( \frac{1}{2}\mathbbm{1}_{N(N_T+N_R)}^T \mathbf{R}-\mathbbm{1}_{N(N_T+N_R)}^T\right)\mathbf{s} \nonumber \\
&=\frac{1}{4}\mathbf{s}^T \mathbf{R}\mathbf{s}+\mathbf{q}^T \mathbf{s}, \label{ising}
\end{align}
where the constant terms have been removed since they have no effect on the optimization problem considered. The Ising formulation in \eqref{ising} has both linear and quadratic terms, which is not compatible with the CIM optical hardware implementations; we note that the considered CIM architecture requires only quadratic terms \cite{ABH1}. We introduce an auxiliary spin variable $s_\alpha$ to convert the linear terms in \eqref{ising} to quadratic terms \cite{ABH2}. Specifically, by introducing the extended vector $\mathbf{s}_0^T=[s_{\alpha}\; \mathbf{s}^T]$, the constraint in \eqref{ising} can be written as
\begin{align}
P_0(\mathbf{s}_0)&=\frac{1}{4}\mathbf{s}_0^T\mathbf{R}_0\mathbf{s}_0+\mathbf{s}_0
\mathbf{C} \mathbf{s}_0=\mathbf{s}_0^T \left(\frac{1}{4}\mathbf{R}_0+\mathbf{C}   \right)\mathbf{s}_0 \nonumber \\
&=\mathbf{s}_0^T \mathbf{J}_0\mathbf{s}_0, \label{const_f}
\end{align}
where the symmetric matrices $\mathbf{R}_0$ and $\mathbf{C}$ are defined as follows
\begin{equation}
\mathbf{R}_{0}=g(\mathbf{R})=\left[ \begin{array}{c|c}
0  & \begin{matrix} 0 & \cdots & 0 \end{matrix} \\ \hline
\begin{matrix} 0 \\ \vdots \\ 0 \end{matrix} & \mathbf{R}
\end{array}\right],
\end{equation}
\begin{equation}
\mathbf{C}=f(\mathbf{q})=\left[
\begin{array}{c}
\begin{matrix}
0 & \frac{1}{2}\mathbf{q}\\				
\frac{1}{2}\mathbf{q}^T & \mathbf{0}_{N(N_{T}+N_{R})\times N(N_{T}+N_{R})}
\end{matrix} \\
\end{array}\right].
\end{equation}
The last mathematical step for the calibration of the constraint in \eqref{const_f} is to set all the diagonal elements of the matrix $\mathbf{J}_0$ to zero and then normalize the resulting matrix such as all its entries take values in the range $[-1,+1]$. In particular, these operations can be represented by the transformation $F_n(\mathbf{J}_0)=\zd(\mathbf{J}_0)/\eta(\zd(\mathbf{J}_0))$ with $\eta(\mathbf{U})\equiv \|\vect(\mathbf{U}) \|_{\infty}$. Therefore, the constraint in \eqref{const_f} is converted into the following CIM-compatible form
\begin{align}
P_0(\mathbf{s}_0)=\mathbf{s}_0^T F_n(\mathbf{J}_0) \mathbf{s}_0. \label{agg}
\end{align}
By using similar analytical steps, we also convert the quadratic binary objective function in \eqref{opt22} into a spin quadratic form without linear terms and normalized quadratic matrix. Specifically, the objective function can be converted as follows 
\begin{align}
\mathbf{b}^T \mathbf{Q} \mathbf{b}\rightarrow \mathbf{s}_0^T F_n(\mathbf{J})\mathbf{s}_0, \label{objf}
\end{align} 
where $\mathbf{J}=\frac{1}{4}g(\mathbf{Q})+f \left(\frac{1}{2}\mathbbm{1}_{N(N_T+N_R)}^T\mathbf{Q} \right)$.

Since CIMs can not handle constraints directly, the final step of the mathematical framework is to combine the objective function in \eqref{objf} with the aggregate constraint in \eqref{agg} by using a penalty scalar $\lambda \in [0,\; 1]$ to ensure the validity of the constraints \cite{YAR}. Since both the objective function and the aggregate constraint are quadratic and normalized, the considered optimization (maximization) problem can take the following final quadratic unconstrained CIM form
\begin{align}
(\overline{s}_{\alpha},\overline{\mathbf{s}})=\arg \max_{s_{\alpha}, \mathbf{s}}\;&(1-\lambda)\mathbf{s}_0^T F_n(\mathbf{J})\mathbf{s}_0-\lambda \mathbf{s}_0^T F_n(\mathbf{J}_0)\mathbf{s}_0 \nonumber \\
&=\mathbf{s}_0^T \Big((1-\lambda)F_n(\mathbf{J})-\lambda F_n(\mathbf{J}_0)  \Big) \mathbf{s}_0.
\end{align}
The above (auxiliary) formulation can be solved in the considered CIM architecture and the produced solution can be used to solve the initial formulation with the equation $\mathbf{\hat{s}}=\overline{s}_{\alpha}\times \overline{\mathbf{s}}$ \cite{ABH1}. It is worth noting that if the CIM solver does not result in any feasible solution, a random selection algorithm is applied without loss of generality. 

It is obvious that the penalty parameter $\lambda$ is critical for the performance of the CIM algorithm; a larger $\lambda$ enforces feasibility (satisfaction of the constraints) but, on the other hand, less resolution in the objective function and vice-versa. In our numerical studies, the impact of the parameter $\lambda$ is demonstrated while its optimal value is adjusted empirically through experimentation.

\subsubsection{Complexity}
Our proposed formulation uses one spin variable to represent the selection decision at each antenna; $N$ states for each of the $N_T + N_R$ antennas leads to a total of $N(N_T + N_R)$ spin variables for representing the problem. One auxiliary spin variable is used to convert all the linear terms into quadratic terms \cite{ABH2}. It takes $\mathcal{O}(N^2(N_T + N_R)^2)$ operations to compute the Ising coefficients corresponding to the SNR maximization objective function ($\mathbf{J}$) and $\mathcal{O}(N^2(N_T + N_R)^3)$ operations to compute the Ising coefficients corresponding to the constraint satisfaction ($\mathbf{J}_0$), leading to a total complexity of $\mathcal{O}(N^2(N_T + N_R)^3)$ for computing the Ising formulation.

\subsection{QA solver}
Next, we solve the considered combinatorial optimization problem, by using the QA solver and by embedding the QA-based formulation into a real-world quantum processing unit, {\it i.e.,} D-Wave Advantage. To apply QA, firstly, the problem at hand is converted to a quadratic unconstrained binary optimization (QUBO) form with $N(N_T+N_R)$ binary variables. By using a similar methodology with Sec. \ref{sec_cim}, the required QUBO formulation is given by
\begin{align}
\arg \max_{\mathbf{b}}\; \mathbf{b}^T \Big(-(1-\lambda)\mathbf{\Theta}+\lambda\mathbf{\Xi} \Big)\mathbf{b},
\end{align}
where the matrix $\mathbf{\Theta}=\mathbf{Q}/\eta(\mathbf{Q})$ captures the objective function, $\mathbf{\Xi}=(\mathbf{R}-2\mathbf{I})/\eta(\mathbf{R}-2\mathbf{I})$ incorporates the antenna configuration constraints, $\lambda$ is the penalty scalar (similar to CIM), and the matrices $\mathbf{Q}$ and $\mathbf{R}$ are given in \eqref{Q_matrix} and \eqref{quad_const}, respectively; the parameter $\lambda$ is adjusted experimentally. It is worth noting that since the QA and the CIM are implemented differently, the dynamics governing each solver evolve in a unique way towards the solution. This implies that the optimal penalty parameters of the two solvers are distinct. Also, the QUBO objective function does not have an inherent minus sign and therefore the combination of the two components is opposite in comparison to the CIM formulation.

The QUBO formulation determines a logical (fully connected) graph, where each node represents a variable and each edge denotes the interaction/correlation between pairs of variables. However, the QA hardware is characterized by sparse graph topologies, where multiple (radio-frequency superconducting) qubits are arranged in unit cells and are interconnected with a limited number of qubits \cite{MCG}. For example, in this work, we consider the D-Wave Advantage-system4.1 architecture that consists of $5,627$ qubits adopting a Pegasus graph topology, where unit cells consist of $8$ qubits and each qubit is inter-connected to up to $15$ qubits \cite{DWAVE}. The process to map the QUBO logical graph into QA hardware graph/topology is called minor-embedding and for practical problems mainly requires the creation of logical qubits where two or more physical/hardware qubits are chained together to act as a single qubit. The intra-chain coupling is a critical system parameter, called chain strength $\overline{J}_F$, which depends on the specific problem under consideration. Determining the optimal value of $\overline{J}_F$ is a non-trivial task and thus it is adjusted experimentally. In case that the final values of the coupled logical qubits are different (after annealing), we have a {\it broken chain} and the final value of the qubit is decided by majority voting. In this work, we adopt the heuristic algorithm {\it minorminer} for the minor-embedding process, which is included in the standard D-Wave Ocean software development kit \cite{DWAVE}. It is worth noting that the embedded QUBO coefficients are represented in a finite precision and range, while the practical annealing operation suffers from analog machine noise (called {\it integrated control error}) due to background susceptibility, qubit flux-noise, quantization, {\it etc.}, which significantly deteriorates the QA performance. To overcome these limitations, multiple anneals are mainly considered to ensure a low-energy solution. An informative overview of the D-Wave architectures and the associated implementation process is given in \cite{MCG,YAR,SRI}.

\section{Capacity Maximization with Heuristic Solvers}\label{ca_so}
\begin{algorithm}[t]
\caption{SA-based Selection}\label{SA_algo}\vspace{1mm}
\hspace*{\algorithmicindent} \textbf{Input} Control parameter $\tau$, scaling factor $\alpha$, termination \\\hspace*{\algorithmicindent} threshold $\epsilon$, channel matrix $\mathbf{G}$
\begin{algorithmic}[1]
\State Select a random configuration $s$ with matrix $\mathbf{H}$ from $\mathbf{G}$
\State Evaluate $C = \log\det(\mathbf{I} + \frac{P}{N_T} \mathbf{H}^H \mathbf{H})$
\Repeat
\State Set $l = 0$
\For{$k = 1 ~\mathrm{to}~ K$}
\State Select a candidate neighbour configuration $s'$ of $s$
\State Let $\mathbf{H}'$ be the channel matrix of configuration $s'$
\State Evaluate $C' = \log\det(\mathbf{I} + \frac{P}{N_T} \mathbf{H}'^H \mathbf{H}')$
\State Let $\Delta C = C' - C$
\If{$\mathrm{random}(0,1) < \min(1,\exp(\Delta C/\tau))$}
\State Update configuration $s$ with $s'$
\State Set $C = C'$ and $l = l + 1$
\EndIf    
\EndFor
\State Set $\delta = l/K$ and $\tau = \alpha \tau$
\Until{$\delta < \epsilon$}
\end{algorithmic}
\end{algorithm}

In this section, we study the optimal antenna configuration that maximizes the end-to-end Shannon capacity. By assuming a symmetric power allocation, the optimization problem ($\textrm{P2}$) is simplified to 
\begin{subequations}\label{opt44}
\begin{align}
\mathbf{(\text{P4})}\;&\max_{\mathbf{X},\mathbf{Y}}\;C_0\equiv \log\det \left(\mathbf{I}+\frac{P}{N_T}\mathbf{X}\mathbf{G}^H\mathbf{Y}\mathbf{G}\mathbf{X}\right) \\
&\textrm{subject to}\;\eqref{c1}, \eqref{c2}.
\end{align}
\end{subequations}
In this case, the objective function cannot be converted to an equivalent Ising problem (quadratic form) and therefore CIM/QA cannot be used directly. As the problem is NP-hard, we consider several heuristic techniques that solve the problem efficiently by keeping the complexity low. In what follows, a discussion for each heuristic is presented.

\subsection{SA-based Selection}
We first propose an SA-based selection algorithm for the considered reconfigurable antenna MIMO system. The main steps of the SA algorithm are given in Algorithm \ref{SA_algo}. Note that the function $\mathrm{random}(0,1)$ draws a random number between $0$ and $1$ from a uniform distribution.

Firstly, define the neighbourhood of a configuration $s$, as the set of all other configurations with $N_T + N_R -1$ antenna states in common with $s$. Moreover, let $\tau$ be the control parameter ({\it i.e.,} the temperature) of the annealing process. The algorithm starts with a random antenna configuration, say $s$, achieving capacity equal to $C$; we refer to $s$ as the \textit{selected configuration}. Then, at each iteration, we pick a \textit{candidate configuration} $s'$ within the neighbourhood of $s$. Let $\Delta C = C' - C$ be the difference between the capacities of the two configurations. If $\Delta C > 0$, that is, if $s'$ achieves a higher capacity than $s$, then accept it ({\it i.e.}, make $s'$ the selected configuration). Otherwise, accept $s'$ with a certain probability, which is a function of the control parameter $\tau$. A common choice for this probability is the \textit{Metropolis criterion} given by $\exp(\Delta C/\tau)$ \cite{SA}. Therefore, acceptance of $s'$ from $s$ occurs with probability
\begin{align}\label{prob1}
p(s,s') = \min(1,\exp(\Delta C/\tau)).
\end{align}
It is clear from \eqref{prob1} that as $\tau$ decreases, the probability of accepting a configuration that achieves a lower capacity becomes smaller. Now, let $\delta$ be the ratio of the number of accepted configurations over the number of candidate configurations, during a period of a specific number of steps $K$. Essentially, a very small $\delta$ means that very few configurations are being accepted, which implies that most likely the algorithm has converged. Therefore, if $\delta$ is below a given threshold, the algorithm is terminated. Otherwise, the control parameter is decreased by a fixed factor $\alpha$, and the process is repeated.

In theory, SA obtains the system's global optimum almost surely with an infinite number of iterations and a slow decrease rate of the control parameter \cite{SA}. In other words, SA is asymptotically an optimization algorithm. However, in practice, SA simulations need to be of finite length. Consequently, there is a possibility that the algorithm may be trapped at a local maximum, from which it becomes unlikely to break away from as the control parameter is decreased. Therefore, in the following sub-section, we consider a parallel tempering algorithm that overcomes this limitation.

\subsection{Parallel Tempering}
Parallel tempering, or replica exchange, simulates $R$ replicas of the annealing process in parallel, each with a different control parameter. Now, a replica with a large control parameter, accepts candidate configurations with higher probability. In other words, this replica samples configurations from a larger sample space. On the other hand, a replica with a small control parameter is restricted to sampling from a more local region. The main idea of parallel tempering is to have replicas exchange their control parameters. In this way, replicas with low control parameters that may potentially be trapped at a local maximum can break away through such exchange. Let $r_i$ denote the $i$-th replica with control parameter $\tau_i$ and capacity of its current selected configuration $C_i$, $i \in \{1,\dots,R\}$. Then, after a specific number of iterations, two replicas $r_i$ and $r_j$ exchange their parameters with probability \cite{JAM2}
\begin{align}\label{pt_prob}
p(i,j) = \min(1,\exp(\Delta\tau [C_j-C_i])),
\end{align}
where $\Delta\tau = 1/\tau_i - 1/\tau_j$. As such, two replicas always exchange parameters, when the one with the higher control parameter has a configuration with a larger capacity.

\subsection{Decoupled Selection with Sequential Elimination}
This sub-optimal heuristic scheme is inspired by conventional antenna selection techniques and sequentially applies row and column elimination in the complete matrix $\mathbf{G}$ such that the capacity loss in each elimination phase is minimum \cite{GOR}. To mathematically express the elimination process, let $\mathbf{g}_k$ denote the $k^{\textrm{th}}$ row of matrix $\mathbf{G}$ and $\mathbf{G}_k$ denote the matrix build on $\mathbf{G}$ after the elimination of the row $\mathbf{g}_k$. The proposed scheme is based on the following fundamental expression 
\begin{align}
&C(\mathbf{G}_k) = \log \det \left (\mathbf{I}+\frac{P}{N_T}\mathbf{G}_k\mathbf{G}_k^H \right) \nonumber \\
&= C(\mathbf{G})+\log \left(1-\frac{P}{N_T}\mathbf{g}_k\left[\mathbf{I}+\frac{P}{N_T}\mathbf{G}^H\mathbf{G}\right]^{-1}\mathbf{g}_k^H \right).
\end{align}
It is obvious that eliminating the $k^{\textrm{th}}$ row results in a capacity reduction, which is captured by the second term in the above expression. Therefore, the optimal row elimination is the one that minimizes the capacity loss, which can be written as
\begin{align}
\hat{k} = \arg \min_k \mathbf{g}_k \left[\mathbf{I} + \frac{P}{N_T}\mathbf{G}^H \mathbf{G}\right]^{-1}\mathbf{g}_k^H.\label{min_loss}
\end{align}

Algorithm \ref{SE_algo} provides the main steps of this method. In the first phase of the algorithm, $N-1$ rows of $\mathbf{G}$ are sequentially eliminated, from each block of rows $(i-1)N+1$ to $iN$ with $i=1,\ldots,N_R$. Thus, the complete matrix $\mathbf{G}$ is converted to a matrix $\overline{\mathbf{G}}$ of dimension $N_R\times N N_T$. The remaining rows correspond to a configuration selection at the receiver side. The second phase of the algorithm consists of the sequential elimination of the columns, that is, configuration selection at the transmitter side. By considering the matrix $\overline{\mathbf{G}}^H$ as the input for this phase, the column elimination process is translated to row elimination. As such, the above scheme can be re-applied accordingly. The final output of the algorithm is a conventional MIMO matrix of dimension $N_R \times N_T$. The algorithm requires $(N_T+N_R) \sum_{k=2}^N k = (N_T + N_R) \left(\frac{N[N+1]}{2}-1\right)$ basic iterations, where a matrix inversion is applied at each iteration; note that the inversion of an $n\times n$ matrix has complexity $\mathcal{O}(n^3)$. Thus, the complexity of the algorithm is $\mathcal{O}((N_T^4+N_R^4)N^5)$.

\begin{algorithm}[t]
\caption{Decoupled Selection with SE}\label{SE_algo}\vspace{1mm}
\hspace*{\algorithmicindent} \textbf{Input} Channel matrix $\mathbf{G}$
\begin{algorithmic}[1]
\State Let $\overline{\mathbf{G}} = \mathbf{G}$
\For{$i = 1 ~\mathrm{to}~ N_R$}
  \For{$j = 1 ~\mathrm{to}~ N-1$}
    \State Find $\hat{k}$ from \eqref{min_loss} for $(i-1)N+1 \leq k \leq iN+1-j$
    \State Eliminate row $\hat{k}$ from $\overline{\mathbf{G}}$
  \EndFor
\EndFor
\State Repeat steps $2-7$ for $\overline{\mathbf{G}} = \overline{\mathbf{G}}^H$ with $N_R = N_T$
\end{algorithmic}
\end{algorithm}

\subsection{SNR-based Maximization Scheme}
The next scheme aims to maximize the SNR objective function. As such, this scheme is sub-optimal in terms of the capacity maximization problem. Specifically, we consider the following low-SNR MIMO approximation \cite[Sec. 8.2.2]{TSE}
\begin{align}
\arg \max_{\mathbf{X},\mathbf{Y}} C \approx \arg \max_{\mathbf{X},\mathbf{Y}} \textrm{Tr}\left( \mathbf{X}\mathbf{G}^H\mathbf{Y}\mathbf{G} \right),\label{snr_max}
\end{align}
which is based on the approximation $\log(1+x) \approx x \log e$ (for small $x$) and thus is efficient for the low-SNR regime. This approximation allows the conversion of the initial capacity maximization problem to an SNR maximization problem. Therefore, the CIM and the QA solvers that have been developed in Sec. \ref{SNR_max} can be applied accordingly. However, it is worth noting that this scheme is highly sub-optimal for intermediate to high SNR values. 

\subsection{Decoupled Selection with ES}
This scheme is based on ES but decouples the selection process between the transmitter and the receiver \cite{GOR2}. The algorithm selects the configuration $\mathbf{Y}^*$ at the receiver that maximizes the determinant, that is,
\begin{align}
\mathbf{Y}^* = \arg \max_{\mathbf{Y}} \det\left(\mathbf{I}+\frac{P}{N_T} \mathbf{G}^H \mathbf{Y} \mathbf{G}\right).
\end{align}
Once $\mathbf{Y}^*$ has been obtained, the algorithm is repeated for all possible configurations $\mathbf{X}$ at the transmitter side, and selects $\mathbf{X}^*$ such that
\begin{align}
\mathbf{X}^* = \arg \max_{\mathbf{X}} \det\left(\mathbf{I}+\frac{P}{N_T} \mathbf{X} \mathbf{G}^H \mathbf{Y} ^*\mathbf{G}\right).
\end{align}
Note that due to the problem's symmetry, one could obtain $\mathbf{X}^*$ first and then $\mathbf{Y}^*$. This decoupling reduces the complexity to $\mathcal{O}(N^{N_T} + N^{N_R})$ compared to $\mathcal{O}(N^{N_T+N_R})$ of the ES.

\section{Evaluation}\label{numerical_st}

Computer simulations and experimental results (in QA D-Wave) are carried-out to evaluate the performance of the proposed heuristic solvers. 

\subsection{Evaluation for CIMs}

We evaluate the performance of our proposed method and benchmark it against ES, NSA and RS schemes. We focus on three key evaluation metrics
\begin{itemize}
    \item $E_\rho \triangleq \mathbbm{E}_{\mathbf{G}}(\mathbbm{E}_{Q_0}(Q_0|\mathbf{G}))$ is the expected value of the SNR-maximization objective function over $10^3$ different channel instances (independent Rayleigh fading channels).
    \item $P_c$ is average probability (over $10^4$ independent Rayleigh fading channels) that CIM generated a solution satisfying the problem constraint. 
    \item $P_{\textrm{oc}}$ is the occurrence probability of distinct solutions ranked in descending order of $Q_0$ over $10^4$ independent  Rayleigh fading channels and $100$ anneals per problem instance (leading to a sample set size of $10^6$).
\end{itemize}

As noted before, CIMs can get stuck in local minima and therefore, it is a common practice to solve the same problem instance multiple times, and each of these runs is referred to as an \textit{anneal} \cite{ABH2}. In this work, we use $1,000$ anneals ($N_a = 1,000$) per problem instance and evaluate both the average performance across all anneals, labeled as CIM (Avg), as well as the performance of	the best solution found by the CIM model, labeled as CIM (Best). The average performance is used to characterize the quality of solutions provided by the CIM solver.

\begin{figure}
\includegraphics[width=\linewidth]{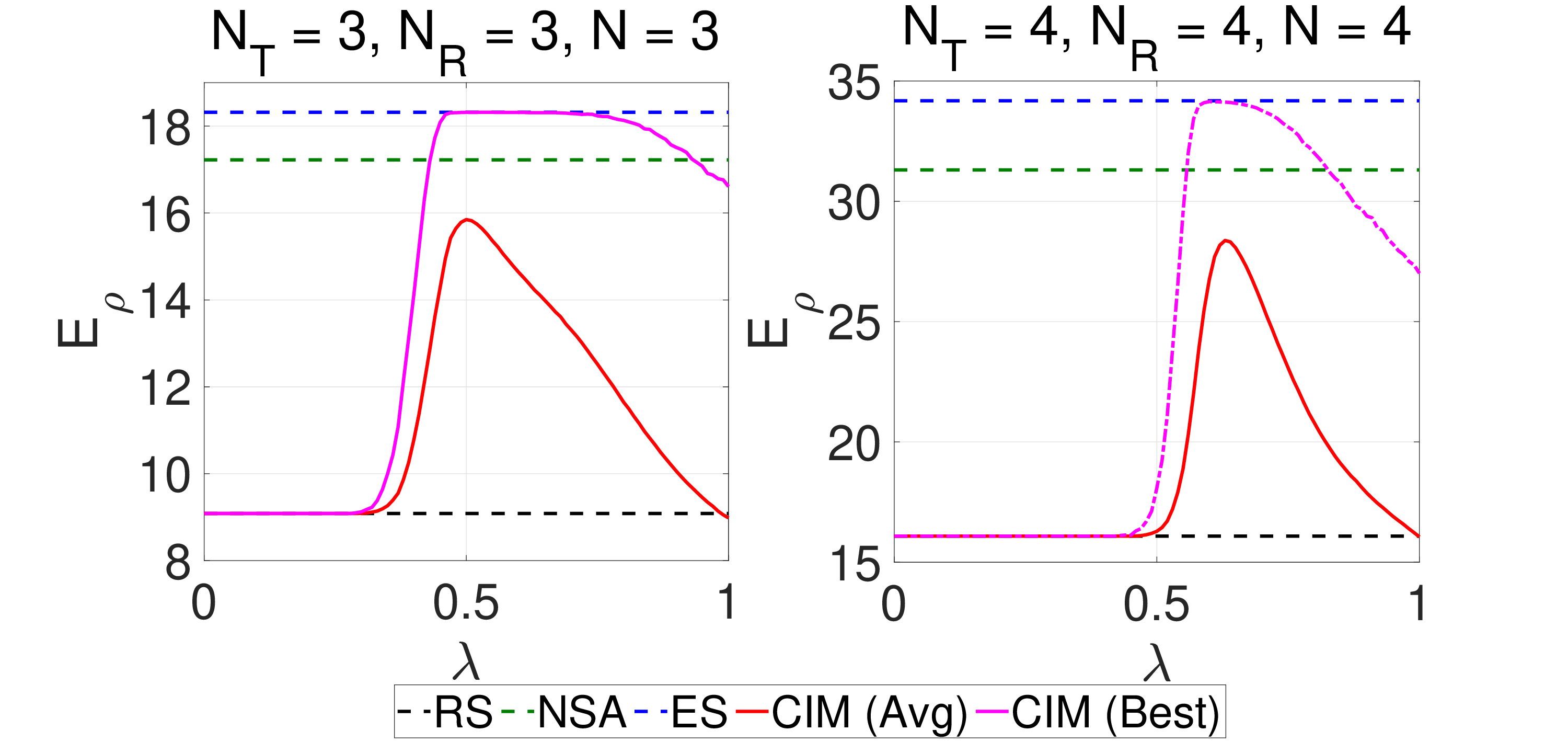}
\caption{[CIM] Expected value of the SNR-maximization objective function with CIM-based antenna configuration selection for different values of the penalty parameter $\lambda$.}\label{fig:snr_v_lambda}
\end{figure}

\begin{figure}
\includegraphics[width=\linewidth]{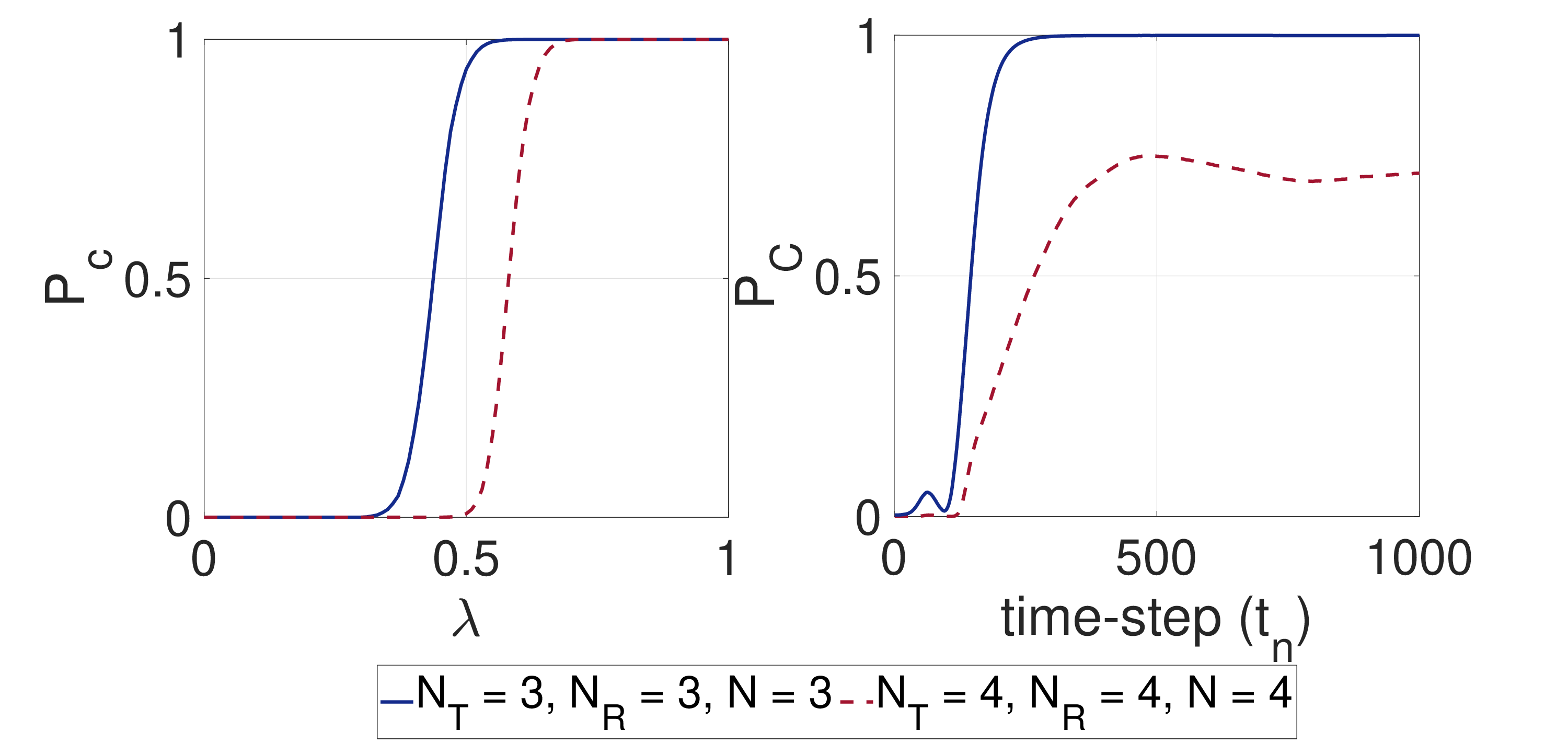}
\caption{[CIM] (left) Probability of constraint satisfaction by CIM-based antenna configuration selection solutions for different values of the penalty parameter $\lambda$. (right) Probability of constraint satisfaction by CIM-based antenna configuration selection solutions as CIM dynamics evolve with time.}\label{fig:prob_v_lambda}
\end{figure}

\subsubsection{Varying the penalty parameter $\lambda$}
We vary the penalty parameter $\lambda$ and observe the performance of our design. Recall that $\lambda$ describes the relative weight given to the constraints of the antenna selection problem, where $\lambda = 1$ corresponds to selecting a valid antenna configuration while ignoring the objective function, and $\lambda = 0$ corresponds to optimizing the objective function while completely ignoring the constraints. In Fig. \ref{fig:snr_v_lambda}, we simulate two MIMO settings ($N_T = 3, N_R = 3, N = 3$) and ($N_T = 4, N_R = 4, N = 4$). We observe that when $\lambda$ is small, the performance of CIM is similar to the RS. This can be explained by the fact that when $\lambda$ is small, the Ising problem is not able to capture the constraints well, and therefore, the CIM returns invalid solutions (Fig. \ref{fig:prob_v_lambda} (left)) and the algorithm defaults to RS. When $\lambda$ is very high, the constraints satisfaction dominates the SNR maximization, and therefore the CIM solutions are valid but perform worse than ES and NSA. In fact, the average CIM performance, \textit{i.e.}, CIM (Avg), performs similar to the RS, as expected. However, at intermediate values of $\lambda$, CIM-based antenna selection can outperform both NSA and RS, and it is possible to empirically tune $\lambda$ to get the best performance.

\begin{figure}
\includegraphics[width=\linewidth]{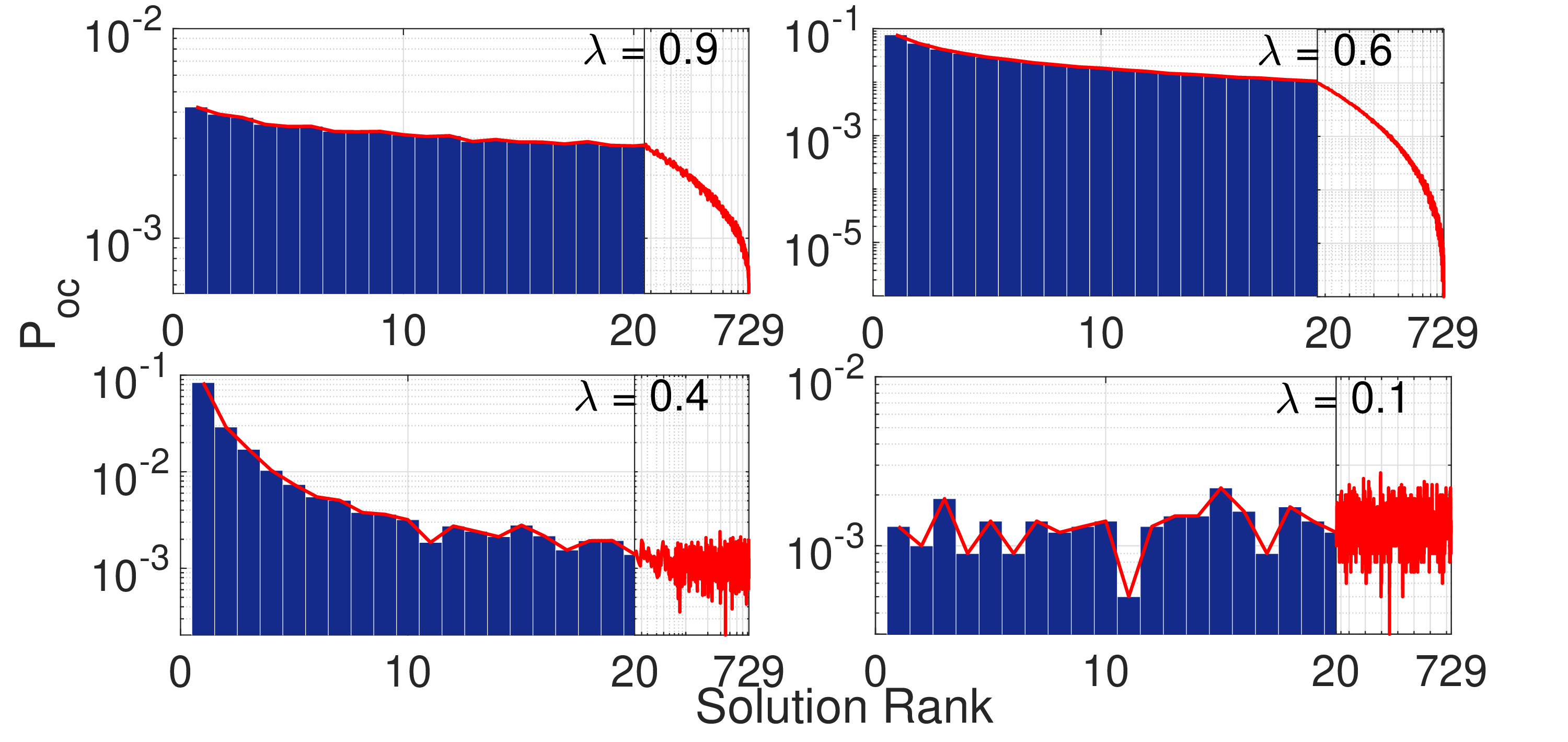}
\caption{[CIM] Occurrence probability of different solutions ranked in descending order of $Q_0$ for $N_T = 3, N_R = 3, N = 3$ over $10^4$ Rayleigh fading channel instances and $100$ anneals per instance.}\label{fig:prob_v_rank}
\end{figure}

\begin{figure}
\includegraphics[width=\linewidth]{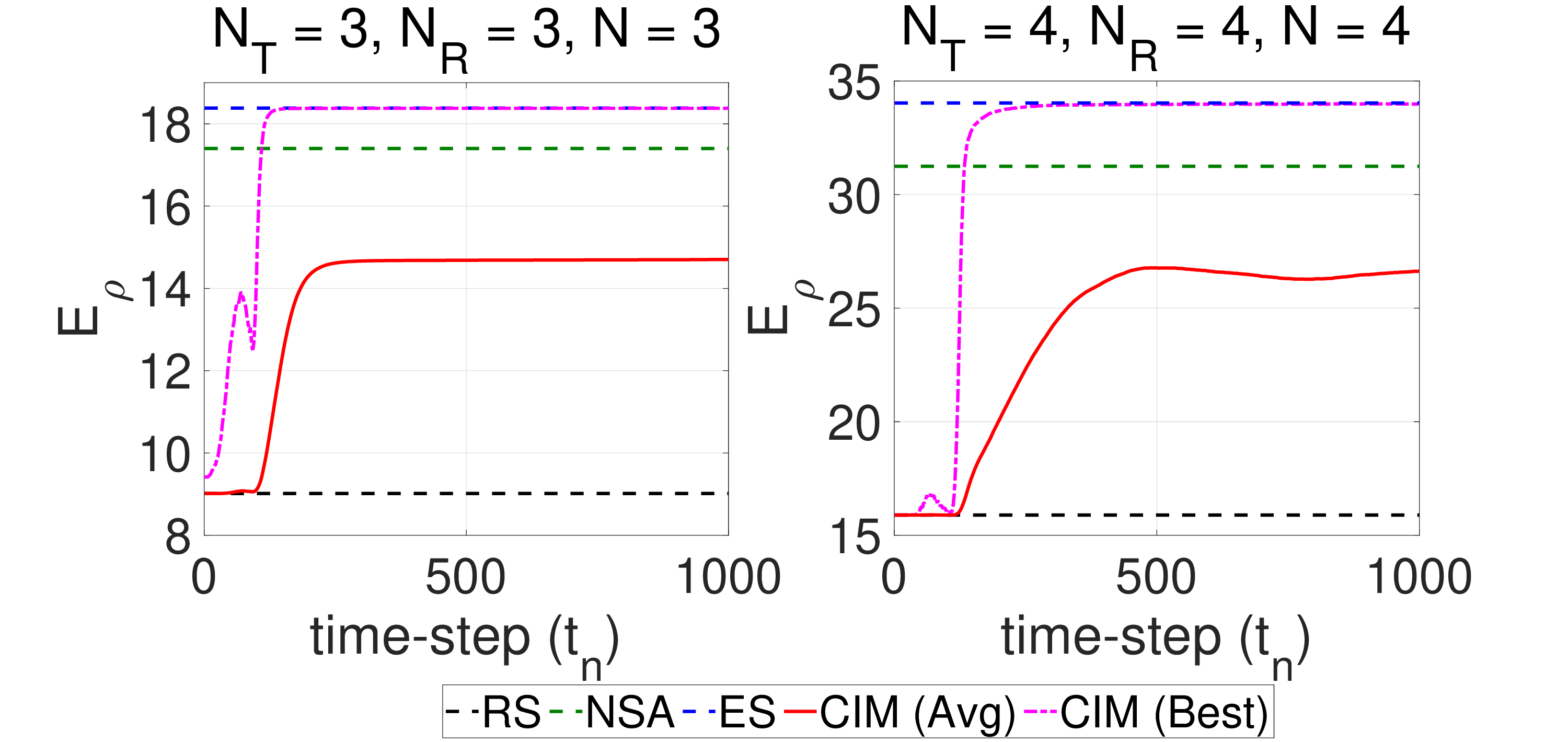}
\caption{[CIM] Expected value of the SNR-maximization objective function with CIM-based antenna configuration selection as CIM dynamics evolves with time.}\label{fig:snr_v_time}
\end{figure}

In Fig. \ref{fig:prob_v_rank}, we plot the occurrence probability of different solutions ranked in descending order of $Q_0$. We see that for $\lambda = 0.1$, given the Ising problem does not capture constraint maximization well, the algorithm resorts to randomly selecting a valid configuration, and $P_{\textrm{oc}}$ is equivalent to a random selection. As $\lambda$ increases to $0.4$, the Ising formulation is able to capture the constraints better and generate valid solutions that optimize the SNR; as a result, we can see that $P_{\textrm{oc}}$ reduces with increasing solution rank and the randomness due to generation of invalid solutions is significantly reduced and seems to be limited to only solutions with high rank. At $\lambda = 0.6$, we see a smooth trend and $P_{\textrm{oc}}$ decreases with increasing solution rank, indicating that the Ising problem is able to properly capture both the SNR maximization objective and the constraints, and CIM outputs valid solutions in accordance with the SNR maximization objective. When $\lambda$ becomes very large, the objective function will be dominated by the constraints, as a result for $\lambda = 0.9$, while $P_{\textrm{oc}}$ still reduces with increasing solution rank, $P_{\textrm{oc}}$ starts to tend towards a uniform distribution among valid solutions. 

\subsubsection{Time-evolution of the CIM solutions}
As noted before, we simulate the behavior of the AHC-based CIM \cite{ahc} by numerical integration of its dynamical equations. Here, we select the optimal $\lambda$ based on the empirical analysis in Fig. \ref{fig:snr_v_lambda} and demonstrate how different performance metrics evolve with time (as we simulate the dynamics of the CIM). Specifically, in Fig. \ref{fig:snr_v_time}, we plot the expectation of the objective function at each step of the numerical integration of the CIM dynamics. We note that, as CIM dynamics evolve with time, its internal states represent a much better solution that progressively improves the objective function. A similar observation holds for the probability of constraint satisfaction. From Fig. \ref{fig:prob_v_lambda} (right), we note that the internal state of CIM becomes increasingly likely to satisfy the constraints of the problem as CIM dynamics evolve with time. Furthermore, we see from Fig. \ref{fig:prob_v_lambda} (right) and Fig. \ref{fig:snr_v_time}, that both $P_c$ and $E_\rho$ appear to reach a steady state after approximately $500$ time-steps, indicating that we need to run the CIM only for that duration. For instance, for the scenario $N_T=N_R=4$ with $N=4$, the ES requires $4^8=65,536$ computations, while CIM (best solution) achieves the optimal configurations in $1,000$ anneals with $500$ time-steps/anneal.   

\subsection{Evaluation for QA D-Wave Advantage}

We now provide experimental results for a state-of-the-art QA device, {\it i.e.,} D-Wave Advantage-system4.1 \cite{ADV}. For each combinatorial problem (channel realization), we consider $N_a=2000$ anneals (unless otherwise defined) with $1$ $\mu\textrm{sec}$ annealing time, where each anneal returns distinct solutions according to a (unknown) probability mass function due to the probabilistic nature of the QA process. We rank the solutions in descending order of their $Q_0$ values as $R_1>R_2>\cdots >R_{N_s}$, where $N_s$ is the number of distinct feasible solutions. The ferromagnetic coupling parameter is equal to $\overline{J}_F=3$. We study the performance in terms of $Q_0$ for each $R_i$ solution as well as the associated occurrence probability $P_{\textrm{oc}}$.

\begin{figure}
\includegraphics[width=\linewidth]{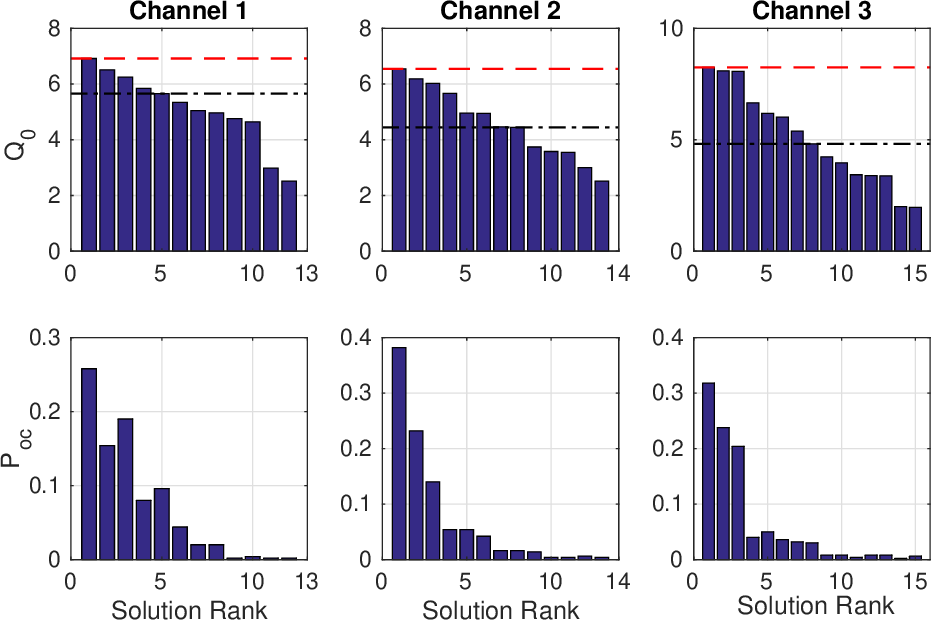}
\caption{[D-Wave] (top) $Q_0$ for the ordered feasible solutions for three channel realizations; ES (dashed line) and NSA (dash-dotted line) are used as benchmarks. (bottom) Probability of occurrence for the ordered feasible solutions. Setup: $\lambda=0.8$, $N_T=N_R=2$ and $N=2$.}
\label{fig:dwave_res1}
\end{figure}

In Fig. \ref{fig:dwave_res1}, we consider three channel instances for a simple MIMO setup with $N_T=N_R=2$ and $N=2$; for this case, $500$ anneals is sufficient to receive efficient results. We plot $Q_0$ and $P_{\textrm{oc}}$ for all feasible returned solutions. It can be seen that the number of feasible distinct solutions $N_s$ varies from $12$ to $15$ while  for all cases, $R_1$ achieves the optimal configuration (similar to ES). Regarding the probability of occurrence, $R_1$ is returned with a probability $0.25$ to $0.4$ while for the considered setup, the feasibility probability is near to $P_{\textrm{fs}}\approx 1$ (QA almost always returns a feasible solution). The parameter $\lambda=0.8$ has been adjusted experimentally and provides an efficient balance between the SNR-based objective function and the constraints/feasibility. A very interesting remark is that both $Q_0$ (energy) results and the associated probability distributions follow a decreasing monotonic behaviour; this is in contrast to conventional BER results in ML studies, where energy and the associated probabilities do not follow any specific pattern \cite{JAM, KASI}.

\begin{figure}
\includegraphics[width=\linewidth]{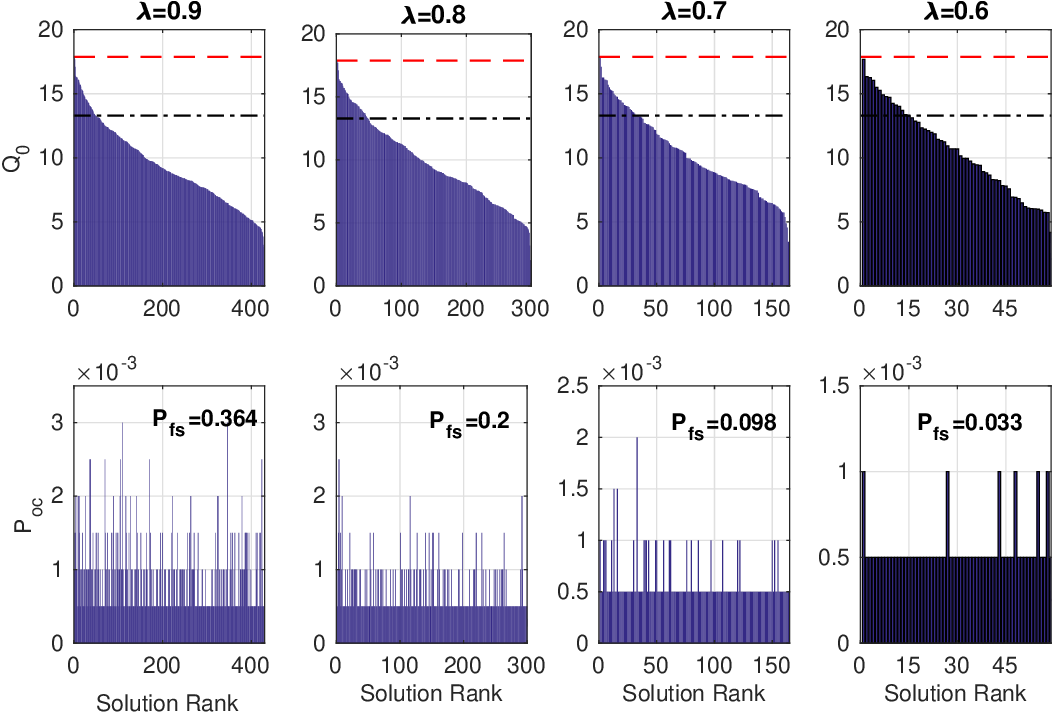}
\caption{[D-Wave] (top) $Q_0$ for the ordered feasible solutions for a single channel realization; ES (dashed line) and NSA (dash-dotted line) are used as benchmarks. (bottom) Probability of occurrence for the ordered feasible solutions. Setup: $N_T=N_R=3$ and $N=3$; $2,000$ anneals.}
\label{fig:dwave_res2}
\end{figure}

\begin{figure}\centering
\includegraphics[width=\linewidth]{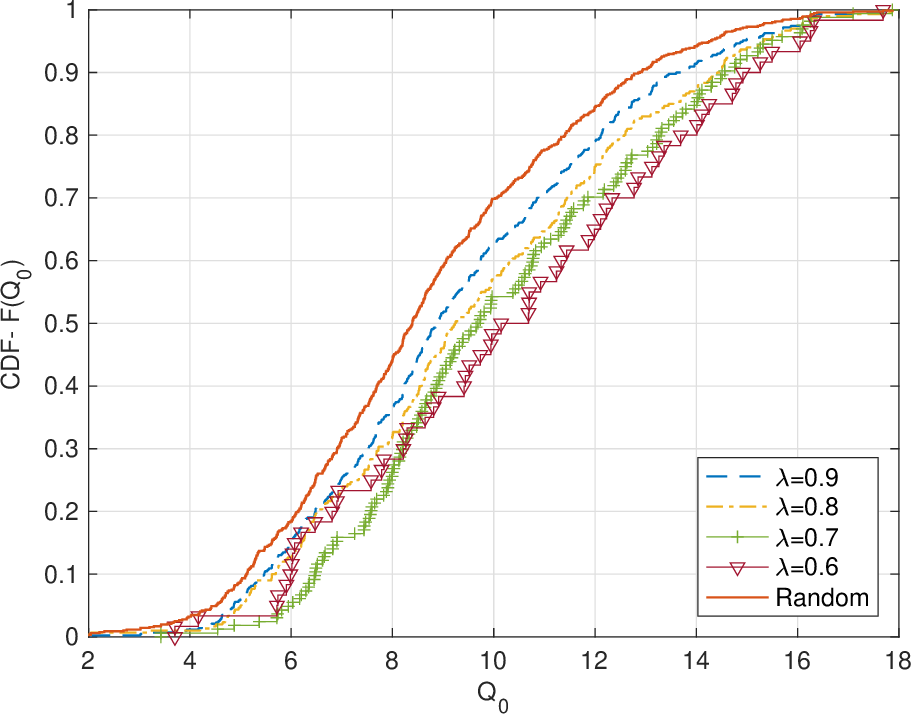}
\caption{[D-Wave] CDF of the D-Wave feasible solutions against random selection policy; $N_T=N_R=3$, $N=3$ and $\lambda \in \{0.9, 0.8, 0.7, 0.6\}$.}
\label{fig:dwave_res3}
\end{figure}

Fig. \ref{fig:dwave_res2} deals with the impact of the parameter $\lambda$ on the QA performance for a more complex MIMO setup with $N_T=N_R=3$ with $N=3$; a single (random) channel realization is considered for this case (similar to a static channel environment). For $\lambda=0.9$, the optimization problem prioritizes the feasibility constraints and therefore the QA returns a feasible solution with probability $P_{\textrm{fs}}=0.368$. As $\lambda$ decreases, the SNR-based objective function is more weighted, and the $Q_0$ performance of the $R_1$ solutions is improving and goes closer to the ES performance. However, if $\lambda$ goes below $0.6$, the performance of the system starts again to decrease. In addition, we observe that, as $\lambda$ decreases, the feasibility probability of the solutions decreases as well (from $0.364$ to $0.033$), which demonstrates an interesting trade-off between feasibility and $Q_0$ performance. To further investigate the statistical behaviour of the D-Wave solutions, in Fig. \ref{fig:dwave_res3}, we plot the CDF of the above (feasible) solutions against the RS policy. We observe that as $\lambda$ increases (which means that the SNR objective is more weighted), the CDF curves shift to the right and the D-Wave significantly outperforms the RS. All these observations demonstrate the critical impact of the parameter $\lambda$ on the QA performance, while $\lambda=0.6$ seems to be an appropriate value for the scenario considered. The feasibility issues which are observed as the MIMO topologies becomes more complex, is an interesting algorithmic design problem for future  work. 

\begin{figure}\centering
\begin{minipage}{0.55\linewidth}\centering
  \includegraphics[width=\linewidth]{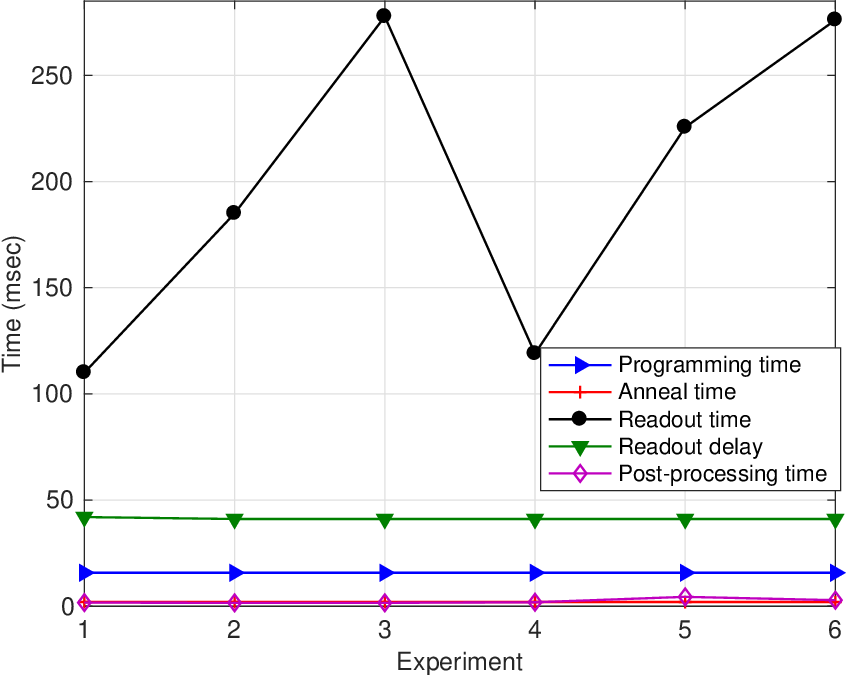}
\end{minipage}%
\begin{minipage}{0.5\linewidth}\centering
  \includegraphics[width=0.7\linewidth]{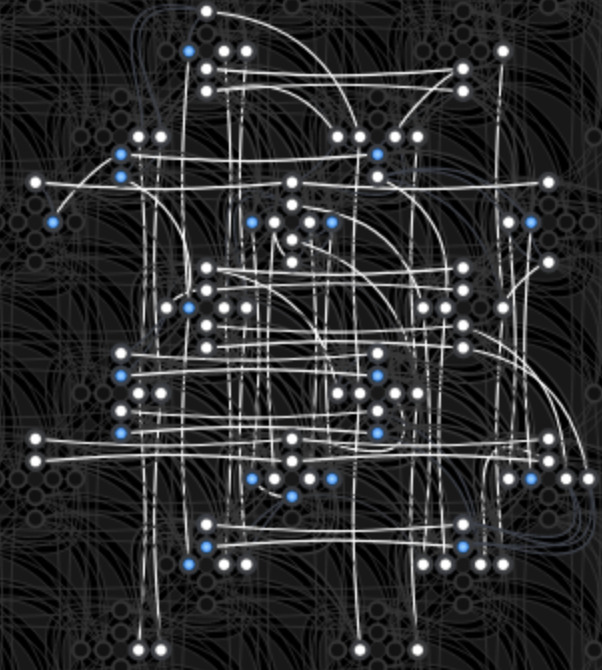}
\end{minipage}
\caption{[D-Wave] (left) Quantum time requirements for various experiments for $2,000$ anneals, (right) Embedding of a QUBO problem into a Pegasus graph (D-Wave Advantage-system4.1), where nodes ($92$ in this case) and edges are qubits and couplers, respectively; $\lambda=0.8$, $N_T=N_R=4$ and $N=4$.}\label{fig:dwave_res4}
\end{figure}

In Fig. \ref{fig:dwave_res4} (left), we deal with the D-Wave time processing performance for a setup with $N_T=N_R=4$, $N=4$ and $\lambda=0.8$; the time results refer to $2,000$ anneals and six independent experiments for the same channel realization. We observe that the total processing time is split into the {\it programming time} (initial pro-processing time to load the QUBO weights onto qubits and coupler biases), the {\it anneal time} (implementation of the QA algorithm, {\it i.e.,} $2$ msec in total), the {\it readout time} (time to read the spin configuration at each anneal), the {\it readout delay} (time to reset the qubits between anneals), and the {\it post-processing time} (time to process the solutions returned by the QA). It can be seen that the anneal time is constant and corresponds to a small fraction ($\approx 1\%$) of the total processing time; the programming time corresponds to almost $6,5\%$ of the total processing, while readout time/readout delay dominate the processing time ($\approx 92,5\%$ of the total time) and seems to be (currently) the main bottleneck for real-time applications. We also observe that although all the timing results are almost constant for all the experiments, the readout time results in large fluctuations, {\it e.g.,} from $100$ msec to $270$ msec; due to the minorminer embedding algorithm, the hardware topology is different in each experiment and this significantly affects the readout time. More sophisticated customized embedding techniques are required in this case to improve processing time performance \cite{KASI}. It is also worth noting that these (overhead) parameters are technology related and they are expected to be significantly smaller in the near future \cite{SRI}.

To visualize the D-Wave Advantage architecture and the mapping/embedding of the QUBO problem into the D-WAVE hardware topology, Fig. \ref{fig:dwave_res4} (right) depicts the Pegasus graph embedding for a QUBO problem with $N_T=N_R=4$, $N=4$, $\lambda=0.8$. We can observe the $2$-D lattice topology and the associated quantum unit cells of the quantum processing unit, as well as the coupling/correlation between the qubits in order to represent the QUBO formulation. 

\subsection{Evaluation for Capacity Maximization}\label{numerical_st2}
Here, we present the performance of the configuration selection algorithms considered in Sec. \ref{ca_so}, with respect to the CDF of the capacity, \textit{i.e.},
\begin{align}
F_C(x) = \mathbb{P}(C < x),
\end{align}
where $C$ is the solution to the optimization problem in \eqref{opt44}. The presented simulation results are obtained with $10^4$ Monte Carlo iterations and consider a normalized value for $P/N_T$ of $10$ dB. The SA-based selection uses the following parameters: $\tau = 0.1$, $\alpha = 0.75$, $\epsilon = 10^{-3}$ and $K = 50N^2N_RN_T$. The parallel tempering algorithm considers two replicas with control parameters $\tau_1 = 0.1$ and $\tau_2 = 0.001$. Each replica runs an annealing process with its control parameter for $200$ steps, following which the two replicas exchange their parameters according to the probability criterion in \eqref{pt_prob}. This is repeated for $200$ iterations, after which the algorithm terminates. We note that the considered parameters were chosen empirically.

\begin{figure}\centering
\includegraphics[width=\linewidth]{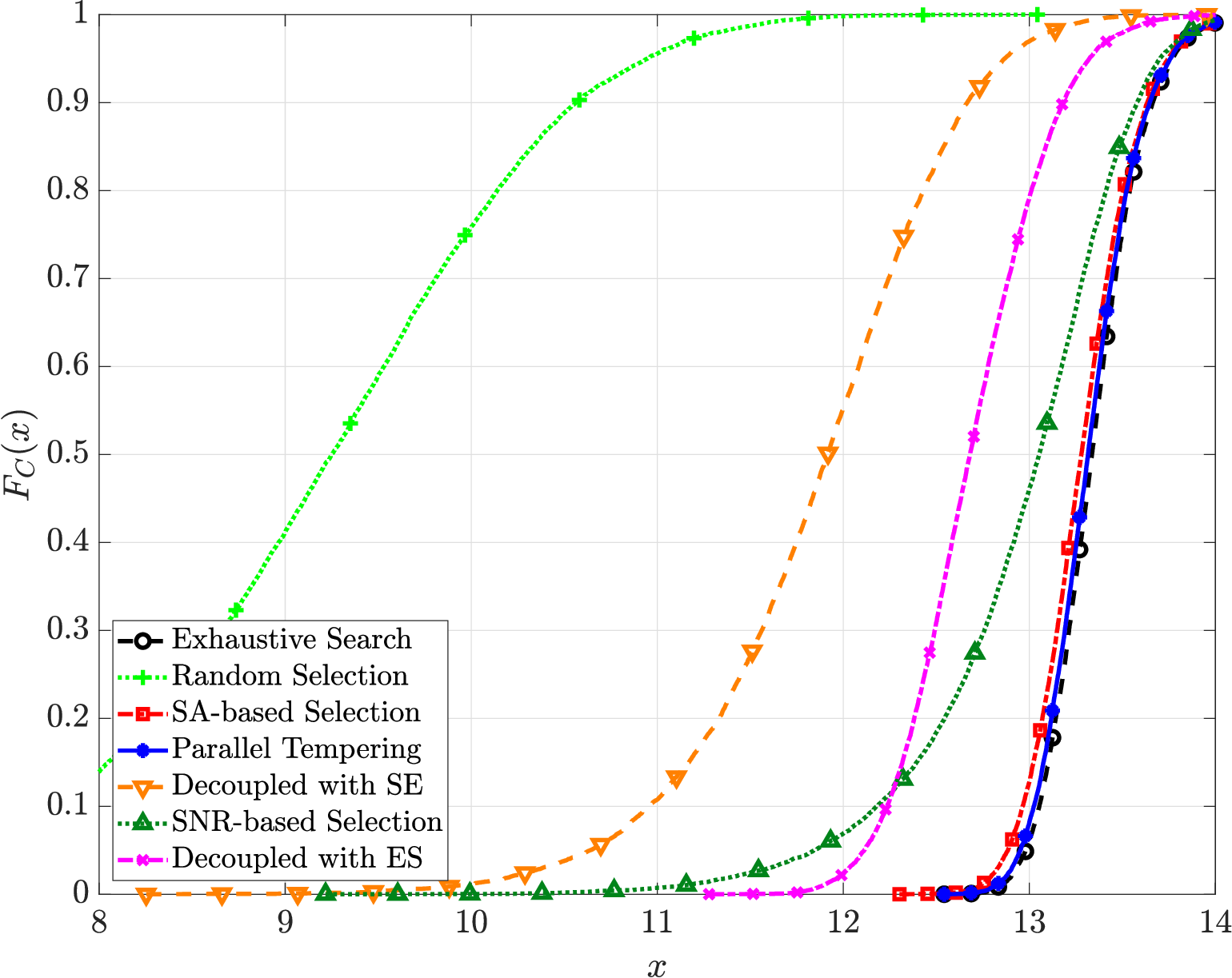}
\caption{Capacity maximization schemes with $N_T = N_R = 3$ and $N = 10$.}\label{CDF_3_3_10}
\end{figure}

\begin{figure}\centering
\includegraphics[width=\linewidth]{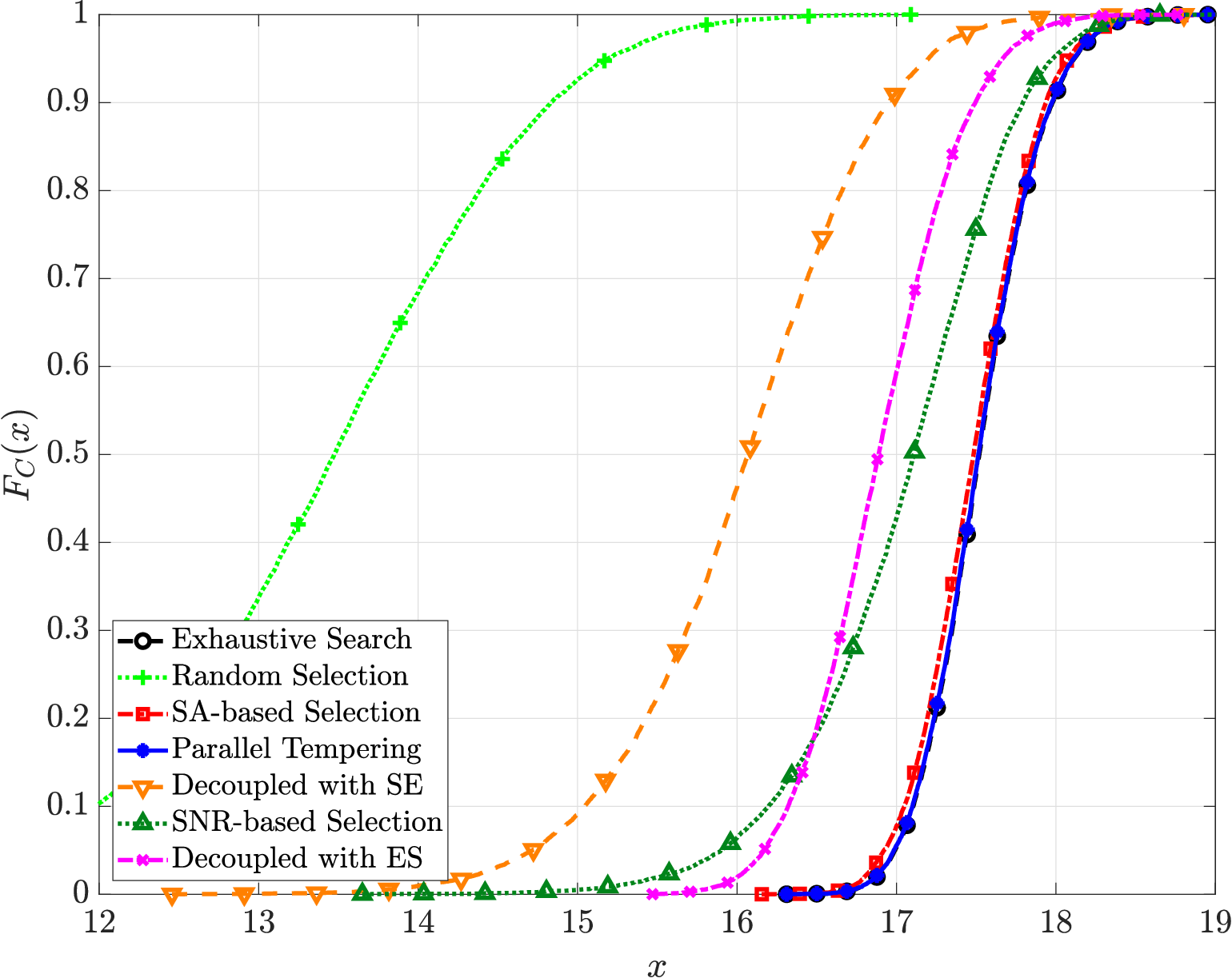}
\caption{Capacity maximization schemes with $N_T = N_R = 4$ and $N = 5$.}\label{CDF_4_4_5}
\end{figure}

We evaluate the performance of the algorithms with two setups: $N_T = N_R = 3$, $N = 10$, illustrated in Fig. \ref{CDF_3_3_10}, and $N_T = N_R = 4$, $N = 5$, depicted in Fig. \ref{CDF_4_4_5}. It is clear that the proposed algorithms (SA-based selection and parallel tempering) achieve near-optimal (in some cases, optimal) performance. The remaining schemes, albeit of lower complexity, are significantly outperformed. Indeed, the performance gap compared to the proposed algorithms increases with the size of the setup. To further compare the considered solvers, we provide in Table \ref{tbl} the average number of configurations each algorithm checks before it terminates with a solution. We can see that, for $N_T = N_R = 3$, $N = 10$, the SA-based selection needs to check $25\%$ of the total number of configurations needed by the ES, whereas parallel tempering needs just $8\%$. For the $N_T = N_R = 4$, $N = 5$ setup, these percentages increase ($32\%$ and $20\%$) since the total number of configurations is smaller. These results clearly demonstrate the effectiveness of the proposed solvers for the configuration selection problem, especially for large MIMO setups. It should also be noted that the number of configurations checked by the SA can be further reduced, with a different set of parameters, by sacrificing some performance gains.

\setlength{\tabcolsep}{10pt}
\renewcommand{\arraystretch}{1.5}
\begin{table}[t]\centering
\caption{Capacity Maximization Solvers \\ Average Number of Configurations Checked}\label{tbl}
\begin{tabular}{|l||*{2}{c|}}\hline
\diagbox[width=4.5cm, height=1cm]{\textbf{Algorithm}}{$N_T\times N_R \times N$} &$3\times 3 \times 10$&$4\times 4 \times 5$\\\hline
ES & $10^6$ & $390,625$\\\hline
SA-based Selection & $248,617$ & $125,927$\\\hline
Parallel Tempering & $80,000$ & $80,000$\\\hline
SNR-based Selection (with CIM) & $1,000$ & $1,000$\\\hline
SNR-based Selection (with QA) & $2,000$ & $2,000$\\\hline
Decoupled Selection with ES & $2,000$ & $1,250$\\\hline
\end{tabular}
\end{table}

\section{Conclusion}\label{concl}
In this paper, we have studied the problem of antenna configuration selection in reconfigurable antenna MIMO systems by using physics-inspired heuristics from classical and quantum mechanics. More specifically, by exploiting the adiabatic time evolution that characterizes the CIM and QA heuristic solvers, the antenna configuration that maximizes the SNR at the receiver is investigated. A rigorous mathematical framework that converts the initial constrained binary combinatorial problem (NP-hard) into unconstrained Ising instances compatible with CIM and QA implementations is developed. The proposed CIM and QA designs are studied for different parameters and performance metrics and we show that they achieve near-optimal performance with polynomial complexity; experimental results in a state-of-the-art QA D-Wave architecture show that the considered problem is appropriate for quantum implementations but also highlight technology limitations for practical real-time applications. In addition, we studied the optimal antenna configuration that maximizes the end-to-end information capacity by using an SA-based algorithm. An improved version of the SA, which exploits parallel tempering to overcome local maxima has also been proposed. Numerical studies show that the SA techniques achieve near-optimal capacity performance and outperform conventional heuristics, which are related (but re-designed here) to the antenna selection literature. An extension of this work is to apply reconfigurable antenna arrays and study the problem of configuration selection in more complex network structures, {\it e.g.,} MU-MIMO systems.  

\bibliographystyle{IEEEtran}
\bibliography{IEEEabrv,ref}

\begin{IEEEbiography}[{\includegraphics[width=1in,height=1.25in,clip,keepaspectratio]{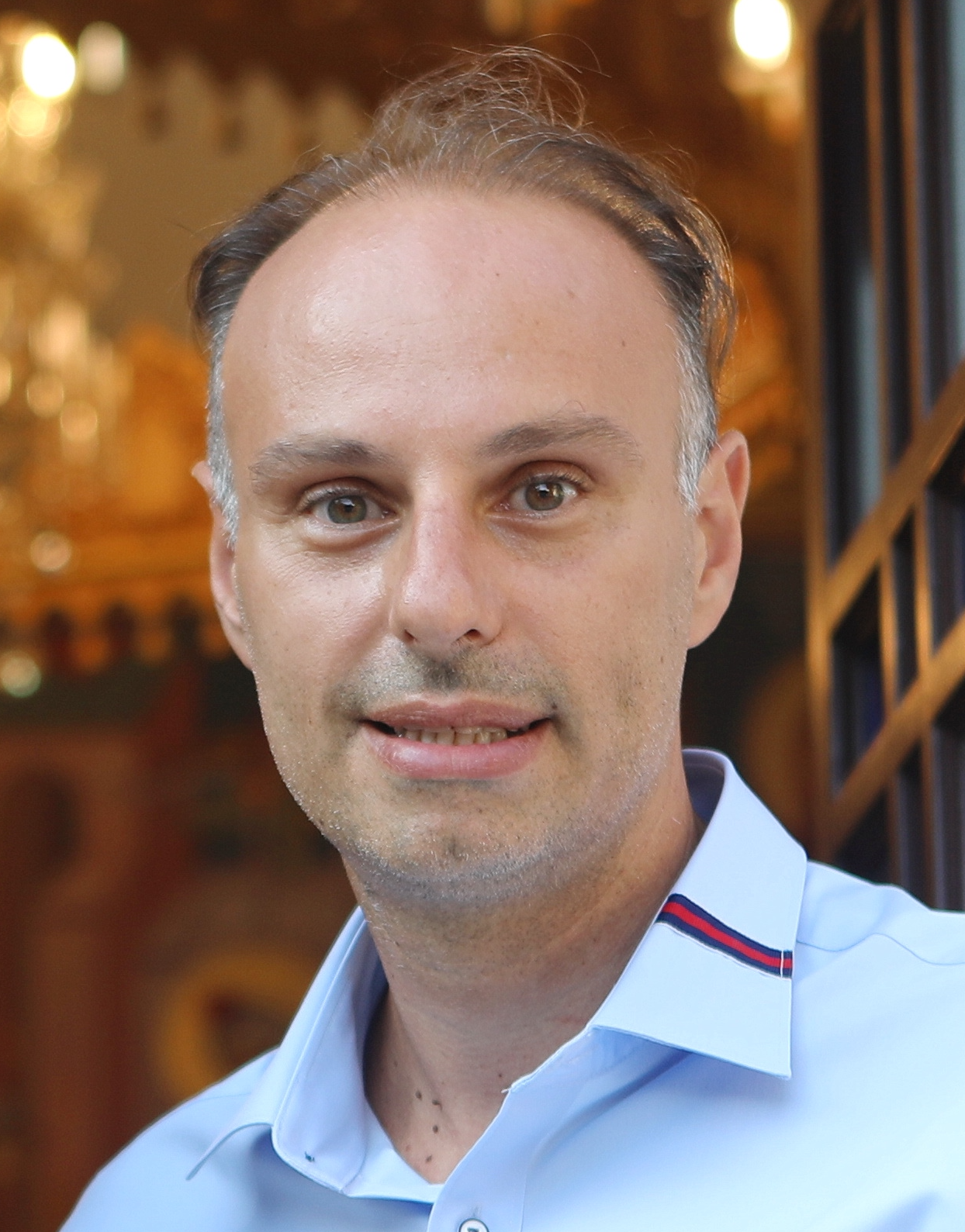}}]{Ioannis Krikidis} (F’19) received the diploma in Computer Engineering from the Computer Engineering and Informatics Department (CEID) of the University of Patras, Greece, in 2000, and the M.Sc and Ph.D degrees from \'Ecole Nationale Sup\'erieure des T\'el\'ecommunications (ENST), Paris, France, in 2001 and 2005, respectively, all in Electrical Engineering. From 2006 to 2007 he worked, as a Post-Doctoral researcher, with ENST, Paris, France, and from 2007 to 2010 he was a Research Fellow in the School of Engineering and Electronics at the University of Edinburgh, Edinburgh, UK. 
	
He is currently a Professor at the Department of Electrical and Computer Engineering, University of Cyprus, Nicosia, Cyprus. His current research interests include wireless communications, quantum computing, 6G communication systems, wireless powered communications, and intelligent reflecting surfaces. Dr. Krikidis serves as an Associate Editor for IEEE Transactions on Wireless Communications, and Senior Editor for IEEE Wireless Communications Letters. He was the recipient of the Young Researcher Award from the Research Promotion Foundation, Cyprus, in 2013, and the recipient of the IEEEComSoc Best Young Professional Award in Academia, 2016, and IEEE Signal Processing Letters best paper award 2019. He has been recognized by the Web of Science as a Highly Cited Researcher for 2017-2021. He has received the prestigious ERC Consolidator Grant for his work on wireless powered communications. 
\end{IEEEbiography}

\begin{IEEEbiography}[{\includegraphics[width=1in,height=1.25in,clip,keepaspectratio]{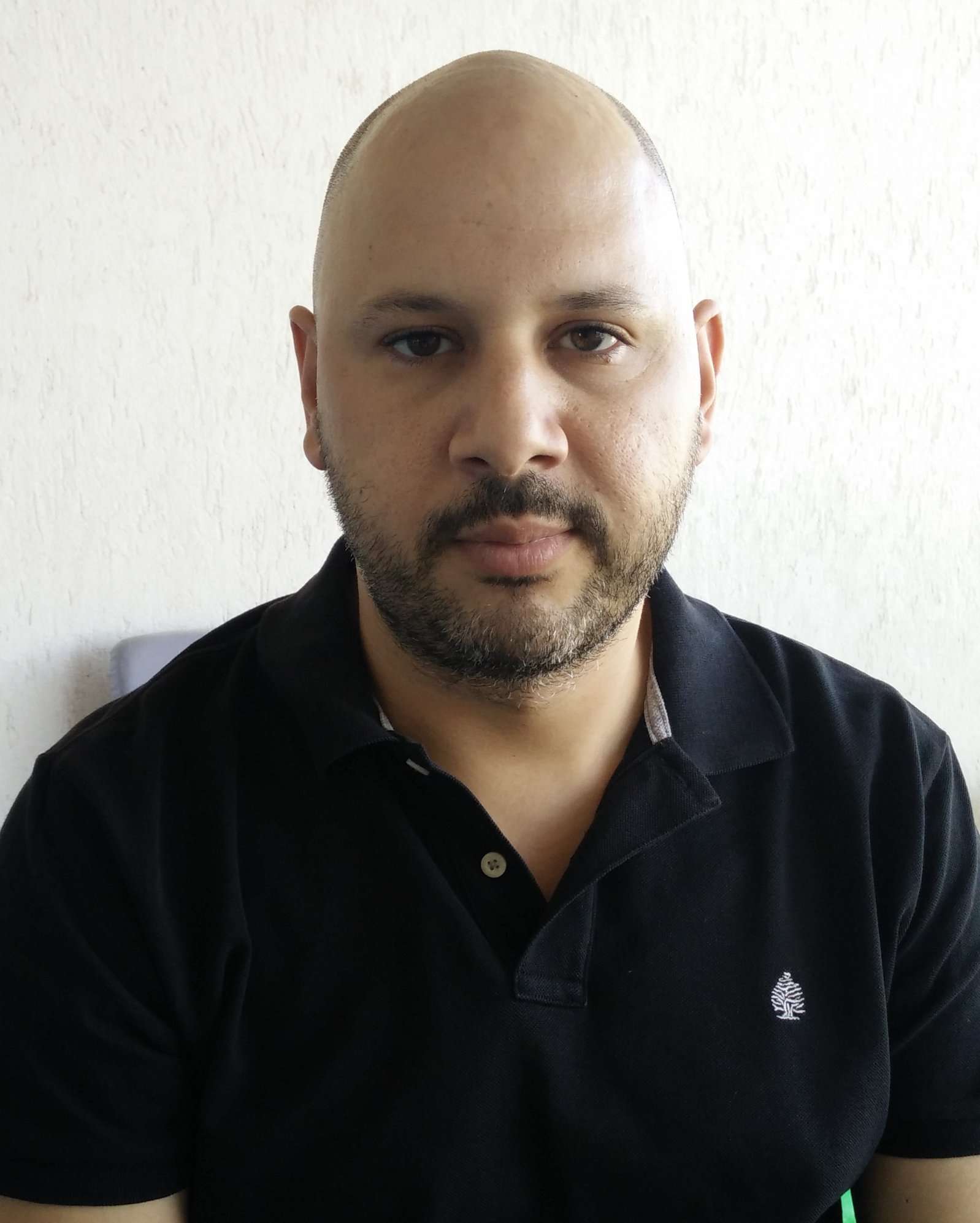}}]{Constantinos Psomas} (M'15–SM'19) holds a B.Sc. in Computer Science and Mathematics (First Class Honours) from Royal Holloway, University of London, an M.Sc. in Applicable Mathematics from the London School of Economics, and a Ph.D. in Mathematics from The Open University, UK. He is currently a Research Fellow with the Department of Electrical and Computer Engineering, University of Cyprus. From 2011 to 2014, he was a Postdoctoral Research Fellow with the Department of Electrical Engineering, Computer Engineering and Informatics, Cyprus University of Technology. His current research interests include wireless powered communications, fluid/movable antennas, and intelligent reflecting surfaces. Dr. Psomas is an Associate Editor for the IEEE Transactions on Communications and the IEEE Wireless Communications Letters. 
\end{IEEEbiography}

\begin{IEEEbiography}[{\includegraphics[width=1in,height=1.25in,clip,keepaspectratio]{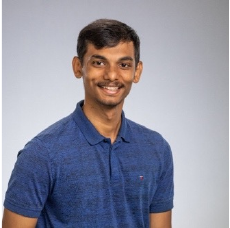}}]{Abhishek Kumar Singh} is a Doctoral Candidate at the Department of Computer Science, Princeton University. His research focusses on classical, quantum, and quantum-inspired computation for next-generation wireless systems. He received the B.Tech. from Indian Institute of Technology, Kanpur (Electrical Engineering, 2016). 
\end{IEEEbiography}

\begin{IEEEbiography}[{\includegraphics[width=1in,height=1.25in,clip,keepaspectratio]{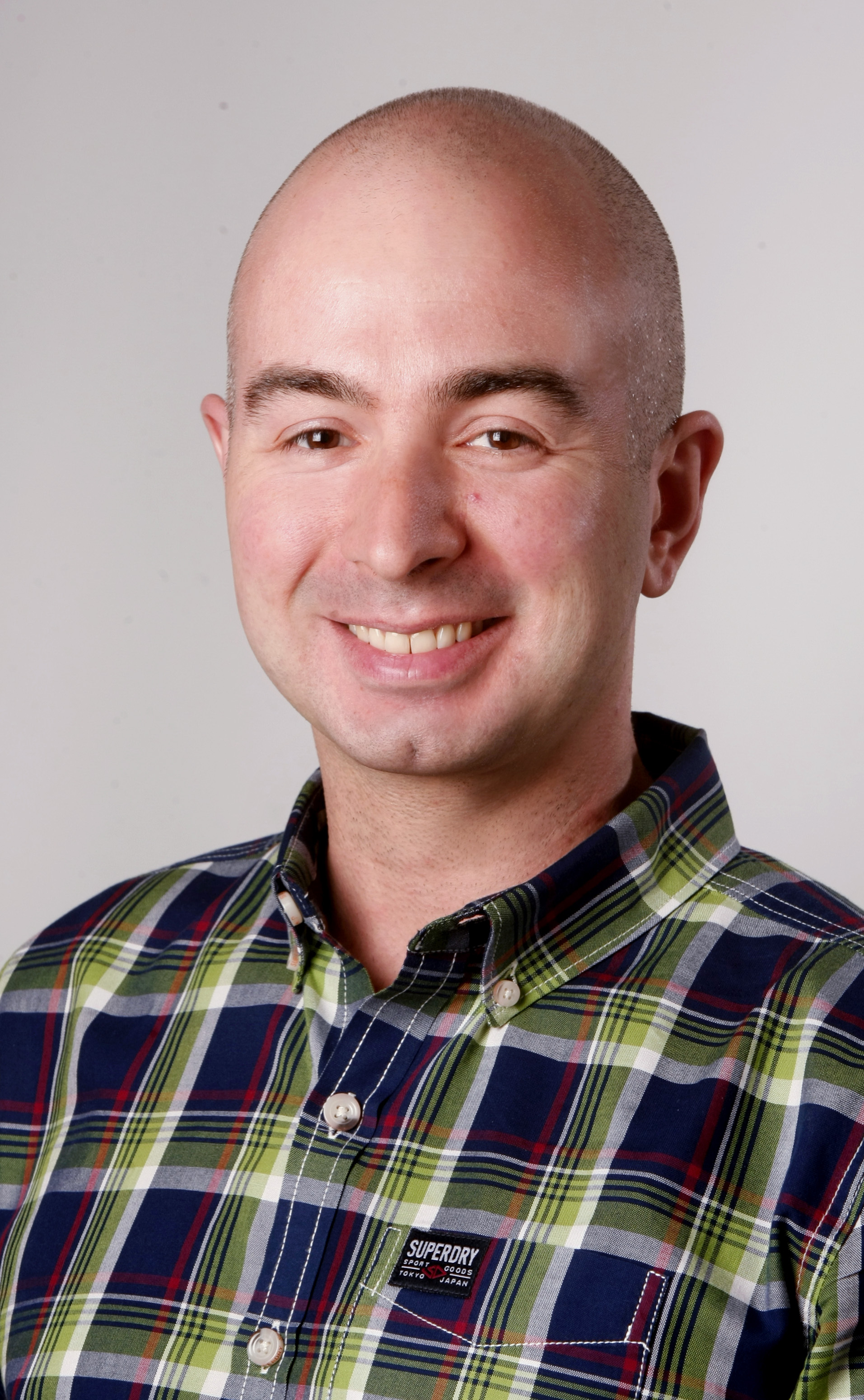}}]{Kyle Jamieson} (Senior Member, IEEE) received the B.S. degree in mathematics and computer science, the M.Eng. degree in computer science and engineering, and the Ph.D. degree in computer science from the Massachusetts Institute of Technology, Cambridge, MA, USA, in 2001, 2002, and 2008, respectively.

He is currently a Professor of Computer Science and Associated Faculty with Electrical and Computer Engineering, Princeton University, Princeton, NJ, USA. His research interests include mobile and wireless systems for sensing, localization, and communication, and on massively-parallel classical, quantum, and quantum-inspired computational structures for NextG wireless communications systems.

Dr. Jamieson was the recipient of the Starting Investigator Fellowship from the European Research Council, a Google Faculty Research Award, and the ACM SIGMOBILE Early Career Award. He was an Associate Editor of IEEE Transactions on Networking from 2018 to 2020 and is a Senior Editor of the IEEE Journal of Selected Areas in Communications. He is a Distinguished Member of the ACM.
\end{IEEEbiography}

\end{document}